\documentclass[a4paper,twoside,twocolumn]{IEEEtran}

\usepackage{amsmath,amsfonts,amssymb}
\usepackage{xspace}
\usepackage{graphicx}
\graphicspath{{./}{./gfx/}}
\usepackage{cite}
\makeatletter
\newcommand{\MathFuncName}[1]{{\operator@font #1}}
\newcommand{\MathFunc}[1]{\mathop{\operator@font #1}\nolimits}
\newcommand{\MathFuncWithLimits}[1]{\mathop{\operator@font #1}\limits}
\makeatother

\newcommand{\Tag}[1]{\mathsf{#1}}        % math font for labels and tags
\newcommand{\V}[1]{\boldsymbol{#1}}      % vector
\newcommand{\M}[1]{\mathbf{#1}}          % matrix
\newcommand{\T}{^\mathrm{T}}             % suffix for transpose
\newcommand{\FT}[1]{\hat{#1}}            % Fourier transform
\renewcommand{\Re}{\MathFunc{Re}}        % real part
\renewcommand{\Im}{\MathFunc{Im}}        % imaginary part
\newcommand{\arc}{\MathFunc{arc}}        % lenght of arc
\newcommand{\Var}{\MathFunc{Var}}        % variance
\newcommand{\mathe}{\mathrm{e}}
\newcommand{\mathi}{\mathrm{i}}

\newcommand{\Paren}[1]{\left(#1\right)}

\newcommand{\Brace}[1]{\left\{#1\right\}}

\newcommand{\QuadTerm}[2]{ #2\T\cdot #1\cdot #2}

\newcommand{\norm}[1]{\Vert #1\Vert}
\newcommand{\Norm}[1]{\left\Vert #1\right\Vert}

\newcommand{\abs}[1]{\vert #1\vert}
\newcommand{\Abs}[1]{\left\vert #1\right\vert}

\newcommand{\avg}[1]{\langle #1\rangle}

\newcommand{\expected}[1]{\mathrm{E}\{#1\}}
\newcommand{\Expected}[1]{\mathrm{E}\left\{#1\right\}}
\newcommand{\bigExpected}[1]{\mathrm{E}\bigl\{#1\bigr\}}

\newcommand{\argmin}{\MathFuncWithLimits{arg\,min}}

\newcommand{\Reals}{\mathbb{R}}

\newcommand{\Complexes}{\mathbb{C}}

\newcommand{\bydef}{\stackrel{\mathrm{def}}{=}}

\newcommand{\cf}{\emph{cf.}\xspace}
\newcommand{\eg}{\emph{e.g.}\xspace}
\newcommand{\ie}{\emph{i.e.}\xspace}
\newcommand{\etal}{\emph{et al.}\xspace}
\newcommand{\etc}{\emph{etc.}\xspace}
\newcommand{\wrt}{with respect to\xspace}
\newcommand{\Eq}[1]{Eq.~(\ref{#1})}
\newcommand{\Fig}[1]{Fig.~\ref{#1}}

  \def\clap#1{\hbox to 0pt{\hss#1\hss}}

\makeatletter

\newcommand{\@SymbolBuilderWithLabel}[4]{%
  \begingroup \escapechar\m@ne\xdef\my@tempa{\string#1}\endgroup
  \expandafter\@ifundefined{\my@tempa}{%
    \def\my@tempb{[} % to avoid mismatching below
    \expandafter\edef\csname\my@tempa\endcsname{%
      \noexpand\@ifnextchar\my@tempb%
          {\csname\my@tempa @b\endcsname}%
          {\csname\my@tempa @a\endcsname}%
    }
    \expandafter\def\csname\my@tempa @a\endcsname{%
      {#2}_{#4}%
    }
    \expandafter\def\csname\my@tempa @b\endcsname[##1]{%
      {{#3}^{#4}_{##1}}%
    }
  }{\@latex@error{\noexpand#1is already defined}\@ehc}%
}
\newcommand{\@SymbolBuilder}[3]{% see \@SymbolBuilderWithLabel for comments
  \begingroup \escapechar\m@ne\xdef\my@tempa{\string#1}\endgroup
  \expandafter\@ifundefined{\my@tempa}{%
    \def\my@tempb{[}
    \expandafter\edef\csname\my@tempa\endcsname{%
      \noexpand\@ifnextchar\my@tempb%
          {\csname\my@tempa @b\endcsname}%
          {\csname\my@tempa @a\endcsname}%
    }
    \expandafter\def\csname\my@tempa @a\endcsname{%
      #2%
    }
    \expandafter\def\csname\my@tempa @b\endcsname[##1]{%
      {{#3}_{##1}}%
    }
  }{\@latex@error{\noexpand#1is already defined}\@ehc}%
}

\newcommand{\SymbolWithLabel}[3]{%
  \@SymbolBuilderWithLabel{#1}{#2}{#2}{#3}}
\newcommand{\Symbol}[2]{%
  \@SymbolBuilder{#1}{#2}{#2}}

\newcommand{\VectorSymbolWithLabel}[3]{%
  \@SymbolBuilderWithLabel{#1}{\V{#2}}{#2}{#3}}
\newcommand{\VectorSymbol}[2]{%
  \@SymbolBuilder{#1}{\V{#2}}{#2}}

\newcommand{\MatrixSymbolWithLabel}[3]{%
  \@SymbolBuilderWithLabel{#1}{\M{#2}}{#2}{#3}}
\newcommand{\MatrixSymbol}[2]{%
  \@SymbolBuilder{#1}{\M{#2}}{#2}}

\makeatother

\newcommand{\ComplexVis}{V}
\newcommand{\VisPhase}{\varphi}
\newcommand{\Powerspectrum}{S}
\newcommand{\Bispectrum}{B}
\newcommand{\PhaseClosure}{\beta}
\newcommand{\Freq}{\nu}               % spatial frequency
\newcommand{\VFreq}{\V{\Freq}}
\newcommand{\Dirn}{\theta}            % view direction
\newcommand{\VDirn}{\V{\Dirn}}
\newcommand{\Posn}{r}                 % projected position
\newcommand{\VPosn}{\V{\Posn}}
\newcommand{\uv}{$(u,v)$\xspace}

\newcommand{\Image}{I}
\newcommand{\Param}{x}
\newcommand{\VParam}{\V{\Param}}
\newcommand{\Gain}{g}
\newcommand{\VGain}{\V{\Gain}}

\newcommand{\GainPhase}{\phi}
\newcommand{\GainSquaredModulus}{\rho}
\newcommand{\OPD}{\delta}

\newcommand{\BasisFunc}{b}
\newcommand{\PixelGrid}{\mathcal{G}}
\newcommand{\FreqSet}{\mathcal{L}}  % set of observed spatial frequencies
\newcommand{\BaseSet}{\mathcal{B}}  % set of telescope pairs that contribute
\newcommand{\ApertureList}{\mathcal{A}} % list of aperture indexes
\newcommand{\ExposureList}{\mathcal{E}} % list of exposures
\newcommand{\Baseline}{B}
\newcommand{\MaxBaseline}{\Baseline_\Tag{max}}

\newcommand{\FeasibleSet}{\mathcal{X}}

\newcommand{\Grid}[1]{\stackrel{%
    {\fboxsep0.09em{\vbox{\hbox{\fbox{}}\vskip0pt}}}}{#1}}

\newcommand{\OTF}{G} % Optical Transfer Function

\Symbol{\Wavelength}{\lambda}
\Symbol{\GridWavelength}{\Grid{\Wavelength}}

\newcommand{\DirLetter}{a}
\Symbol{\Dir}{\V{\DirLetter}}
\Symbol{\GridDir}{\Grid{\V{\DirLetter}}}

\newcommand{\ParamLetter}{x}
\VectorSymbol{\X}{\ParamLetter}

\MatrixSymbol{\XForm}{T}
\MatrixSymbol{\OTFop}{\OTF}
\MatrixSymbol{\DFTop}{F}
\MatrixSymbol{\InterpOp}{R}
\MatrixSymbol{\ModelOp}{A}
\MatrixSymbol{\PSFop}{H}
\MatrixSymbol{\ApodizationOp}{S}

\Symbol{\PSF}{h}
\VectorSymbolWithLabel{\Xpsf}{x}{\Tag{psf}}

\newcommand{\DataLetter}{y}
\VectorSymbol{\Data}{\DataLetter}

\VectorSymbol{\Model}{m}

\VectorSymbol{\Error}{e}

\VectorSymbol{\Residual}{r}

\newcommand{\DataTag}{\Tag{data}}
\newcommand{\PriorTag}{\Tag{prior}}
\newcommand{\ErrorTag}{\Tag{err}}
\newcommand{\ModelTag}{\Tag{model}}

\newcommand{\PowerspectrumTag}{\Tag{ps}}
\newcommand{\BispectrumTag}{\Tag{bisp}}
\newcommand{\PhaseClosureTag}{\Tag{cl}}
\newcommand{\BestTag}{+}

\newcommand{\DataLevel}{\eta_\DataTag}
\newcommand{\PriorLevel}{\eta_\PriorTag}

\MatrixSymbolWithLabel{\Cerror}{C}{\ErrorTag}
\MatrixSymbolWithLabel{\Werror}{W}{\ErrorTag}
\MatrixSymbolWithLabel{\Cprior}{C}{\PriorTag}
\MatrixSymbolWithLabel{\Wprior}{W}{\PriorTag}
\MatrixSymbol{\Identity}{I}

\newcommand{\Fcost}{f}

\newcommand{\Fdata}{\Fcost_\DataTag}
\newcommand{\Fprior}{\Fcost_\PriorTag}
\newcommand{\Weight}{w}

\newcommand{\Mira}{MiRA\xspace}
\newcommand{\BSMEM}{BSMEM\xspace}

\newcommand{\Wipe}{\textsc{Wipe}\xspace}
\newcommand{\Wisard}{\textsc{Wisard}\xspace}
\newcommand{\Clean}{\textsc{Clean}\xspace}

\begin{document}

\title{Image Reconstruction in Optical Interferometry}

%\author{Eric~Thi\'ebaut
%  \thanks{%Centre de Recherche Astrophysique de Lyon;
%  Email: thiebaut@obs.univ-lyon1.fr}
%  and
%  Jean-Fran\c{c}ois Giovannelli
%  \thanks{Email: Giova@IMS-Bordeaux.fr}
%}

%\author{\'E.\ Thi\'ebaut and J.-F.\ Giovannelli \thanks{\'E.\ Thi\'ebaut is
%    with Centre de Recherche Astrophysique de Lyon, CNRS/UMR 5574,
%    Observatoire de Lyon, Universit\'e Lyon 1, \'Ecole Normale Sup\'{e}rieure
%    de Lyon, E-mail: \texttt{thiebaut@obs.univ-lyon1.fr}.  J.-F.\ Giovannelli
%    is with Laboratoire Automatique Productique et Signal, Bordeaux, E-mail:
%    \texttt{Giova@IMS-Bordeaux.fr}}}

\author{\'E.\ Thi\'ebaut \thanks{\'E.\ Thi\'ebaut is with Centre de Recherche
    Astrophysique de Lyon, CNRS/UMR 5574, Observatoire de Lyon, Universit\'e
    Lyon 1, \'Ecole Normale Sup\'{e}rieure de Lyon, E-mail:
    \texttt{thiebaut@obs.univ-lyon1.fr}.}  and J.-F.\ Giovannelli
  \thanks{J.-F.\ Giovannelli is with the Laboratoire d'Int\'egration du
    Mat\'eriau au Syst\`eme, \'Equipe Signal-Image, Universit\'e de Bordeaux
    1, 33405 Talence, E-mail: \texttt{Giova@IMS-Bordeaux.fr}.}}

%\author{%
%  \IEEEauthorblockN{Eric~Thi\'ebaut}
%  \IEEEauthorblockA{%
%    Centre de Recherche Astrophysique de Lyon\\
%    CNRS/UMR 5574, Observatoire de Lyon, \\
%    Universit\'e Lyon 1, \'Ecole Normale Sup\'{e}rieure de Lyon,
%    E-mail: thiebaut@obs.univ-lyon1.fr%
%  }
%  \and
%  \IEEEauthorblockN{Jean-Fran\c{c}ois Giovannelli}
%  \IEEEauthorblockA{%
%    Laboratoire Automatique Productique et Signal\\
%    Bordeaux, France.\\
%    E-mail: Giova@IMS-Bordeaux.fr%
%  }
%}

\maketitle

\begin{abstract}
This tutorial paper describes the problem of image reconstruction from
interferometric data with a particular focus on the specific problems
encountered at optical (visible/IR) wavelengths.  The challenging issues in
image reconstruction from interferometric data are introduced in the general
framework of inverse problem approach.  This framework is then used to
describe existing image reconstruction algorithms in radio interferometry and
the new methods specifically developed for optical interferometry.

\emph{Index Terms} -- aperture synthesis, interferometry, image
reconstruction, inverse problems, regularization.
\end{abstract}

\section{Introduction}

Since the first multi-telescope optical interferometer
\cite{Labeyrie-1975-Vega}, considerable technological improvements have been
achieved.  Optical (visible/IR) interferometers are now widely open to the
astronomical community and provide means to obtain unique information from
observed objects at very high angular resolution (sub-milliarcsecond).  There
are numerous astrophysical applications: stellar surfaces, environment of
pre-main sequence or evolved stars, central regions of active galaxies, \etc
See \cite{Quirrenbach-2001-optical_interferometry,
  Monnier-2003-interferometry, Perrin-2009-VLTI_science} for comprehensive
reviews about optical interferometry and recent astrophysical results.  As
interferometers do not directly provide images, reconstruction methods are
needed to fully exploit these instruments.  This paper aims at reviewing image
reconstruction algorithms in astronomical interferometry using a general
framework to formally describe and compare the different methods.

Multi-telescope interferometers provide sparse measurements of the Fourier
transform of the brightness distribution of the observed objects (\cf Section
\ref{sec:interferometric-data}).  Hence the first problem in image
reconstruction from interferometric data is to cope with voids in the sampled
spatial frequencies.  This can be tackled in the framework of inverse problem
approach (\cf Section \ref{sec:image-reconstruction}).  At optical
wavelengths, additional problems arise due to the missing of part of Fourier
phase information, and to the non-linearity of the direct model.  These issues
had led to the development of specific algorithms which can also
be formally described in the same general framework (\cf Section
\ref{sec:optical-interferometry}).

\section{Interferometric Data}
\label{sec:interferometric-data}

%\subsection{Complex Visibilities}
%\label{sec:complex-visibilities}

\begin{figure}[!t]
  \centering
  \includegraphics[width=60mm]{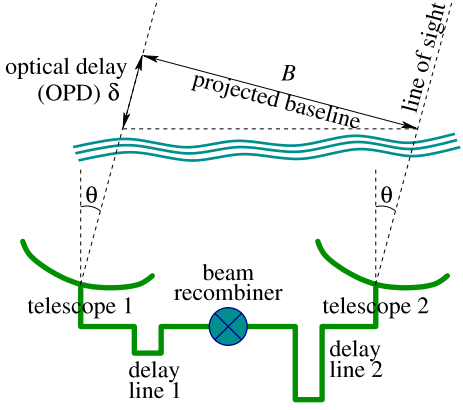}
  \caption{Geometrical layout of an interferometer. $B$ is the projected
    baseline, $\Dirn$ is the view angle and $\OPD$ is the geometrical optical
    path difference which is compensated by the delay lines.}
  \label{fig:layout}
\end{figure}

The instantaneous output of an optical interferometer is the so-called
\emph{complex visibility} $\ComplexVis_{j_1,j_2}(t)$ of the fringes given by
the interferences of the monochromatic light from the $j_1$-th and the
$j_2$-th telescopes at instant $t$~\cite{Monnier-2003-interferometry}:
\begin{equation}
  \label{eq:instantaneous-complex-visibility}
  \ComplexVis_{j_1,j_2}(t) = \Gain_{j_1}(t)^{\star} \, \Gain_{j_2}(t) \,
  \FT{\Image}\Paren{\VFreq_{j_1,j_2}(t)}
\end{equation}
where $\FT{\Image}(\VFreq)$ is the Fourier transform of $\Image(\VDirn)$, the
brightness distribution of the observed object in angular direction $\VDirn$,
$\Gain_{j}(t)$ is the complex amplitude throughput for the light from the
$j$-th telescope and $\VFreq_{j_1,j_2}(t)$ is the spatial frequency sampled by
the pair of telescopes $(j_1,j_2)$ (see Fig.~\ref{fig:layout}):
\begin{equation}
  \label{eq:instantaneous-frequency}
  \VFreq_{j_1,j_2}(t) = \frac{\VPosn_{j_2}(t) - \VPosn_{j_1}(t)}{\Wavelength}
\end{equation}
with $\Wavelength$ the wavelength and $\VPosn_j(t)$ the projected position of
the $j$-th telescope on a plane perpendicular to the line of sight.  These
equations assume that the diameters of the telescopes are much smaller than
their projected separation and that the object is an incoherent light source.
An interferometer therefore provides sparse measurements of the Fourier
transform of the brightness distribution of the observed object.  The top-left
panel of Fig.~\ref{fig:dirty-map} shows an example of the sampling of spatial
frequencies by an interferometer.

%Fig.~\ref{fig:uv-coverage} shows the sampling of spatial frequencies -- the
%so-called \uv coverage -- at different wavelengths for one night of
%observation by IOTA interferometer \cite{Dyck_et_al-1995-IOTA_first_results}
%equipped with IONIC instrument \cite{Berger_et_al-2003-IONIC}.

In practice, the complex visibility is measured during a finite exposure
duration:
\begin{equation}
  \label{eq:measured-complex-visibility}
  \ComplexVis_{j_1,j_2,m}^\DataTag = \avg{\ComplexVis_{j_1,j_2}(t)}_{m}
  + \ComplexVis_{j_1,j_2,m}^\ErrorTag
\end{equation}
where $\avg{\,}_{m}$ denotes averaging during the $m$-th exposure and
$\ComplexVis_{j_1,j_2,m}^\ErrorTag$ stands for the errors due to noise and
modeling approximations.  The exposure duration is short enough to consider
the projected baseline $\VPosn_{j_2}(t) - \VPosn_{j_1}(t)$ as constant, thus:
\begin{equation}
  \label{eq:averaged-complex-visibility}
  \avg{\ComplexVis_{j_1,j_2}(t)}_{m} \simeq
  \OTF_{j_1,j_2,m}\,
  %\avg{\Gain_{j_1}(t)^{\star} \, \Gain_{j_2}(t)}_{m} \,
  \FT{\Image}\Paren{\VFreq_{j_1,j_2,m}}
\end{equation}
with $\VFreq_{j_1,j_2,m} = \avg{\VFreq_{j_1,j_2}(t)}_{m} \simeq
\VFreq_{j_1,j_2}(t_m)$, $t_m=\avg{t}_{m}$ the mean exposure time and
$\OTF_{j_1,j_2,m} = \avg{\Gain_{j_1}(t)^{\star} \, \Gain_{j_2}(t)}_{m}$ the
effective optical transfer function (OTF). The fast variations of the
instantaneous OTF are mainly due to the random optical path differences (OPD)
caused by the atmospheric turbulence.  In long baseline interferometry, two
telescopes are separated by more than the outer scale of the turbulence, hence
their OPDs are independent.  Furthermore, the exposure duration is much longer
than the evolution time of the turbulence (a few $10\,$ms) and averaging can
be approximated by expectation:
\begin{displaymath}
  %\OTF_{j_1,j_2,m} =
  \avg{\Gain_{j_1}(t)^{\star} \, \Gain_{j_2}(t)}_{m}
  \simeq \Expected{\Gain_{j_1}(t)^{\star}}_{m} \,
  \Expected{\Gain_{j_2}(t)}_{m}
\end{displaymath}
with $\expected{}_{m}$ the expectation during the $m$-th exposure.  During
this exposure, the phase of $\Gain_{j}(t)$ is
$\GainPhase_{j}(t)=\GainPhase_{j,m} + (2\,\pi/\lambda)\,\OPD_j(t)$ with
$\GainPhase_{j,m}=\avg{\GainPhase_{j}(t)}_{m}$ the static phase aberration and
$\OPD_j(t)\sim\mathcal{N}(0,\sigma_{\OPD}^2)$ the OPD which is a zero-mean
Gaussian variable with the same standard deviation for all telescopes
\cite{Roddier1981}.  For a given telescope, the amplitude and phase of the
complex throughput can be assumed independent, hence:
\begin{displaymath}
  \Expected{\Gain_{j}(t)}_{m}
  \simeq  \Expected{\abs{\Gain_{j}(t)}}_{m} \, 
  \bigExpected{\mathe^{\mathi\,\GainPhase_{j}(t)}}_{m}
  \simeq \Gain_{j,m} \, \mathe^{-\frac{1}{2}\sigma_{\GainPhase}^2}
\end{displaymath}
with $\Gain_{j,m}=\abs{\Gain_{j}(t_m)}\,\exp(-\mathi\,\GainPhase_{j,m})$ and
$\sigma_{\GainPhase}^2=(2\,\pi/\lambda)^2\,\sigma_{\OPD}^2$ the variance of
the phase during an exposure.  The OTF is finally:
\begin{equation}
  \label{eq:mean-otf}
  \OTF_{j_1,j_2,m} = \avg{\Gain_{j_1}(t)^{\star} \, \Gain_{j_2}(t)}_{m}
  \simeq \Gain_{j_1,m}^{\star} \, \Gain_{j_2,m} \,
  \mathe^{-\sigma_{\GainPhase}^2}\,.
\end{equation}

At long wavelengths (radio), the phase variation during each exposure is
small, hence $\OTF_{j_1,j_2,m} \simeq \Gain_{j_1,m}^{\star} \,
\Gain_{j_2,m}\not=0$.  If some means to calibrate the $\Gain_{j,m}$'s are
available, then image reconstruction amounts to deconvolution (\cf Section
\ref{sec:image-reconstruction}); otherwise, \emph{self-calibration} (\cf
Section~\ref{sec:self-calibration}) has been developed to jointly estimate the
OTF and the brightness distribution of the object given the measured complex
visibilities.

%\subsection{Measurements in Optical Interferometry}
%\label{sec:non-linear-data}

At short wavelengths (optical), the phase variance exceeds a few squared
radians and $\OTF_{j_1,j_2,m}\simeq0$, hence the object's complex visibility
cannot be directly measured.  A first solution would be to compensate for the
OPD errors in real time using fast delay lines.  This solution however
requires a bright reference source in the vicinity of the observed object and
dedicated instrumentation \cite{Delplancke_at_al-2003-Prima} that is currently
in development and not yet available.  An alternative solution consists in
integrating non-linear estimators that are insensitive to telescope-wise phase
errors.  This requires high acquisition rates (about $1000\,\mathrm{Hz}$ in
the near infrared) and involves special data processing but otherwise no
special instrumentation.

To overcome loss in visibility transmission due to fast varying OPD errors,
current optical interferometers integrate the power spectrum (for
$j_1\not=j_2$):
\begin{equation}
  \label{eq:mean-powerspectrum}
  \Powerspectrum_{j_1,j_2,m}
  = \avg{\abs{\ComplexVis_{j_1,j_2}(t)}^2}_m
  \simeq \GainSquaredModulus_{j_1,m} \,
         \GainSquaredModulus_{j_2,m} \,
         \abs{\FT{\Image}(\VFreq_{j_1,j_2,m})}^2 \,,
\end{equation}
with $\GainSquaredModulus_{j,m} = \avg{\abs{\Gain_{j}(t)}^2}_m$ the mean
squared modulus of the complex throughput of the $j$-th telescope during the
$m$-th exposure.  By construction, the $\GainSquaredModulus_{j,m}$'s are
insensitive to the phase errors and so is the power spectrum.  Unlike that of
the complex visibility, the transfer function $\GainSquaredModulus_{j_1,m} \,
\GainSquaredModulus_{j_2,m}$ of the power spectrum is not negligible.  This
transfer function can be estimated by simultaneous photometric calibration
and, to compensate for remaining static effects, from the power spectrum of a
reference source (a so-called \emph{calibrator}).  Hence the object power
spectrum $\abs{\FT{\Image}(\VFreq_{j_1,j_2,m})}^2$ can be measured by
$\Powerspectrum_{j_1,j_2,m}$ in spite of phase errors due to the turbulence.

To obtain Fourier phase information (which is not provided by the
power spectrum), the \emph{bispectrum} of the complex visibilities is measured:
\begin{align}
  \Bispectrum_{j_1,j_2,j_3,m}
  &= \avg{\ComplexVis_{j_1,j_2}(t)\,\ComplexVis_{j_2,j_3}(t)\,
     \ComplexVis_{j_3,j_1}(t)}_m \notag \\
  &\simeq \GainSquaredModulus_{j_1,m} \,
     \GainSquaredModulus_{j_2,m} \,
     \GainSquaredModulus_{j_3,m} \, \notag \\
  & \quad {} \times
     \FT{\Image}(\VFreq_{j_1,j_2,m}) \,
     \FT{\Image}(\VFreq_{j_2,j_3,m}) \,
     \FT{\Image}(\VFreq_{j_3,j_1,m}) \, ,
     \label{eq:mean-bispectrum}
\end{align}
where $j_1$, $j_2$ and $j_3$ denote three different telescopes.  As for the
power spectrum, the transfer function $\GainSquaredModulus_{j_1,m} \,
\GainSquaredModulus_{j_2,m} \, \GainSquaredModulus_{j_3,m}$ of the bispectrum
can be calibrated.  Since this transfer function is real, it has no effect on
the phase of the bispectrum (the so-called \emph{phase closure}) which is
equal to that of the object:
\begin{align}
  \PhaseClosure_{j_1,j_2,j_3,m}
  &\equiv \arg(\Bispectrum_{j_1,j_2,j_3,m}) \notag \\
  &= \arg(\FT{\Image}(\VFreq_{j_1,j_2,m}) \,
          \FT{\Image}(\VFreq_{j_2,j_3,m}) \,
          \FT{\Image}(\VFreq_{j_3,j_1,m}))\, .
  \label{eq:phase-closure}
\end{align}
Some phase information is however missing.  Indeed, from all the interferences
between $T$ telescopes (in a non-redundant configuration), $T\,(T - 1)/2$
different spatial frequencies are sampled but the phase closure only yields
$(T - 1)\,(T - 2)/2$ linearly independent phase
estimates~\cite{Monnier-2003-interferometry}.  The deficiency of phase
information is most critical for a small number of telescopes.  Whatever the
number of telescopes is, at least the information of absolute position of the
observed object is lost.

In practice, obtaining the power spectrum and the bispectrum involves
measuring the instantaneous complex visibilities (that is, for a very short
integration time compared to the evolution of the turbulence) and averaging
their power spectrum and bispectrum over the effective exposure time.  Being
non-linear functions of noisy variables, these quantities are biased but the
biases are easy to remove \cite{DaintyGreenaway1979,Wirnitzer1985}.  To
simplify the description of the algorithms, we will consider that the
\emph{de-biased} and \emph{calibrated} power spectrum and bispectrum are
available as input data for image reconstruction, thus:
\begin{align}
  \Powerspectrum_{j_1,j_2,m}^\DataTag
  &= \abs{\FT{\Image}(\VFreq_{j_1,j_2,m})}^2
     + \Powerspectrum_{j_1,j_2,m}^\ErrorTag \, ,
  \label{eq:powerspectrum-data}
  \\
  \Bispectrum_{j_1,j_2,j_3,m}^\DataTag
  &= \FT{\Image}(\VFreq_{j_1,j_2,m}) \,
     \FT{\Image}(\VFreq_{j_2,j_3,m}) \,
     \FT{\Image}(\VFreq_{j_3,j_1,m}) \notag\\
  & \quad{+}\: \Bispectrum_{j_1,j_2,j_3,m}^\ErrorTag
  \label{eq:bispectrum-data}
\end{align}
where $\Powerspectrum_{j_1,j_2,m}^\ErrorTag$ and
$\Bispectrum_{j_1,j_2,j_3,m}^\ErrorTag$ are zero-mean terms that account for
noise and model errors.
%Note that the spatial frequencies involved in the bispectrum make a
%\emph{closed triangle}:
%\begin{displaymath}
%  \VFreq_{j_1,j_2,m} + \VFreq_{j_2,j_3,m} + \VFreq_{j_3,j_1,m} = 0 \, .
%\end{displaymath}
Instead of the complex bispectrum data, we may consider the phase closure
data:
\begin{align}
  \PhaseClosure_{j_1,j_2,j_3,m}^\DataTag
  &= \arc\bigl(\VisPhase(\VFreq_{j_1,j_2,m}) +
          \VisPhase(\VFreq_{j_2,j_3,m}) \notag\\
   &\hspace{4em}{+}\:\VisPhase(\VFreq_{j_3,j_1,m}) +
        \PhaseClosure_{j_1,j_2,j_3,m}^\ErrorTag\bigr) \, ,
  \label{eq:phase-closure-data}
\end{align}
where $\VisPhase(\VFreq)=\arg(\FT{\Image}(\VFreq))$ is the Fourier phase of
the object brightness distribution, $\arc(\,)$ wraps its argument in the range
$(-\pi,+\pi]$ and $\PhaseClosure_{j_1,j_2,j_3,m}^\ErrorTag$ denotes the
errors.

%%%%%%%%%%%%%%%%%%%%%%%%%%%%%%%%%%%%%%%%%%%%%%%%%%%%%%%%%%%%%%%%%%%%%%%%%%%%%%%
%%%%%%%%%%%%%%%%%%%%%%%%%%%%%%%%%%%%%%%%%%%%%%%%%%%%%%%% IMAGE RECONSTRUCTION %
%%%%%%%%%%%%%%%%%%%%%%%%%%%%%%%%%%%%%%%%%%%%%%%%%%%%%%%%%%%%%%%%%%%%%%%%%%%%%%%

\section{Imaging from Sparse Fourier Data}
\label{sec:image-reconstruction}

We consider here the simplest problem of image reconstruction given sparse
Fourier coefficients (the complex visibilities) and first assuming that the
OTF has been calibrated.

%==============================================================================
%====================================================== DATA AND IMAGE MODELS =
%==============================================================================

\subsection{Data and Image Models}

To simplify the notation, we introduce the \emph{data vector}
$\Data\in\Complexes^L$ which collates all the measurements:
$\Data[\ell]=\ComplexVis_{j_1,j_2,m}^\DataTag$ with $\ell\sim(j_1,j_2,m)$ to
denote a one-to-one mapping between index $\ell$ and triplet $(j_1,j_2,m)$.
Long baseline interferometers provide data for a limited set
$\FreqSet=\{\VFreq_k\}_{k=1,\ldots,K}$ of observed spatial frequencies.  For
each $\VFreq_k$, there is a non-empty set $\BaseSet_k$ of telescope pairs and
exposures such that:
\begin{displaymath}
  (j_1,j_2,m) \in \BaseSet_k
  \quad\Longleftrightarrow\quad
  \VPosn_{j_2,m} - \VPosn_{j_1,m} = \Wavelength\,\VFreq_k
\end{displaymath}
or equivalently:
\begin{equation}
  \label{eq:base-set-def}
  \BaseSet_k \bydef \left\{(j_1,j_2,m) \in \ApertureList^2\times\ExposureList
  %\quad\text{s.t.}\quad
  ;\ \VPosn_{j_2,m} - \VPosn_{j_1,m} = \Wavelength\,\VFreq_k \right\}
\end{equation}
with $\ApertureList$ and $\ExposureList$ the sets of apertures (telescopes or
antennae) and exposure indexes, and $\VPosn_{j,m}=\avg{\VPosn_j(t)}_{m}$ the
mean position of the $j$-th telescope during the $m$-th exposure.  Introducing
$\BaseSet_k$ and the set $\FreqSet$ of observed frequencies is a simple way to
account for all possible cases (with or without redundancies, multiple data
sets, observations from different interferometers, \etc).  Note that, if every
spatial frequency is only observed once, then $L=K$ and we can use $\ell=k$.

The image is a parametrized representation of the object brightness
distribution.  A very general description is given by a linear expansion:
\begin{equation}
  \label{eq:general-image-model}
  \Image(\VDirn) = \sum_{n=1}^{N} \Param_n \, \BasisFunc_n(\VDirn)
  \quad \stackrel{\mathrm{F.T.}}{\longrightarrow}\quad
  \FT{\Image}(\VFreq) = \sum_{n=1}^{N} \Param_n \, \FT{\BasisFunc}_n(\VFreq)
  \,,  
\end{equation}
where $\{\BasisFunc_n(\VDirn)\}_{n=1,\ldots,N}$ are basis functions and
$\VParam\in\Reals^N$ are the image parameters, for instance, the values of the
image \emph{pixels}, or wavelet coefficients.  Given a grid of angular
directions $\PixelGrid=\{\VDirn_n\}_{n=1,\ldots,N}$ and taking
$\BasisFunc_n(\VDirn) = \BasisFunc(\VDirn - \VDirn_n)$, a \emph{grid model} is
obtained:
\begin{align}
  \Image(\VDirn) &= \sum_{n=1}^{N} \Param_n \BasisFunc(\VDirn - \VDirn_n)
  \notag\\
  \label{eq:grid-image-model}
  &\stackrel{\mathrm{F.T.}}{\longrightarrow}\quad
  \FT{\Image}(\VFreq) = \FT{\BasisFunc}(\VFreq)\,
  \sum_{n=1}^{N} \Param_n \,
  \mathrm{e}^{-\mathrm{i}\,2\,\pi\,\VDirn_n\cdot\VFreq} \,.
\end{align}
%\begin{equation}
%  \label{eq:building-block}
%  \BasisFunc_n(\VDirn) = \BasisFunc(\VDirn - \VDirn_n) \, .
%\end{equation}
Using an equispaced grid, the usual pixelized image representation is obtained
with pixel shape $\BasisFunc(\VDirn)$.  The function $\BasisFunc(\VDirn)$ can
also be used as a \emph{building-block} for image reconstruction
\cite{Hofmann_Weigelt-1993-building_blocks}.  Alternatively,
$\BasisFunc(\VDirn)$ may be seen as the \emph{neat beam} that sets the
effective resolution of the image \cite{Lannes_et_al-1997-Clean_and_Wipe}.

The size of the synthesized field of view and the image resolution must be
chosen according to the extension of the observed object and to the resolution
of the interferometer, see \eg \cite{Lannes_et_al-1997-Clean_and_Wipe}.  To
avoid biases and rough approximations caused by the particular image model,
the grid spacing $\Delta\Dirn$ should be well beyond the limit imposed by the
longest baseline:
\begin{equation}
  \label{eq:diffraction-limit}
  \Delta\Dirn \ll \frac{\lambda}{2\,\MaxBaseline}
\end{equation}
where $\MaxBaseline=\max_{j_1,j_2,t}\abs{\VPosn_{j_1}(t) - \VPosn_{j_2}(t)}$
is the maximum projected separation between interfering telescopes.
Oversampling by a factor of at least 2 is usually used and the pixel size is
given by: $\Delta\Dirn \lesssim \lambda/(4\,\MaxBaseline)$.  To avoid aliasing
and image truncation, the field of view must be chosen large enough and
without forgetting that the reciprocal of the width of the field of view also
sets the sampling step of the spatial frequencies.

The model of the complex visibility at the observed spatial frequencies is:
\begin{equation}
  \label{eq:sampled-image-FT}
  \ComplexVis_{k}(\VParam) = \FT{\Image}(\VFreq_k)
  = \sum_{n=1}^{N} \XForm[k,n] \, \Param_{n}
  \,,
\end{equation}
where the coefficients of the matrix $\XForm\in\Complexes^{K\times{}N}$ are
$\XForm[k,n] = \FT{\BasisFunc}_n(\VFreq_k)$ or $\XForm[k,n] =
\FT{\BasisFunc}(\VFreq_k) \, \mathe^{-\mathi\,2\,\pi\,\VDirn_n\cdot\VFreq_k}$
depending which model of \Eq{eq:general-image-model} or
\Eq{eq:grid-image-model} is used.  The matrix $\XForm$ performs the Fourier
transform of non-equispaced data, which is a very costly operation.  This
problem is not specific to interferometry, similar needs in crystallography,
tomography and bio-medical imaging have led to the development of fast
algorithms to approximate this operation
\cite{Potts_et_al-2001-NFFT_tutorial}.  For instance:
\begin{equation}
  \label{eq:Fourier-approximation}
  \XForm \simeq \InterpOp\cdot\DFTop\cdot\ApodizationOp \, ,
\end{equation}
where $\DFTop\in\Complexes^{N\times{}N}$ is the fast Fourier transform (FFT)
operator, $\InterpOp\in\Complexes^{K\times{}N}$ is a linear operator to
interpolate the discrete Fourier transform of the image
$\FT{\VParam}=\DFTop\cdot\VParam$ at the observed spatial frequencies and
$\ApodizationOp$ is diagonal and compensates the field of view apodization (or
spectral smoothing) caused by $\InterpOp$.

In radio astronomy a different technique called \emph{regridding}
\cite{Thompson_Bracewell-1974-Fourier_interpolation,1989ASPC....6..117X} is
generally used, which consists in interpolating the data (not the model) onto
the grid of discrete frequencies.  The advantage is that, when there is a
large number of measurements, the number of data points is reduced, which
speeds up further computations.  There are however a number of drawbacks to
the regridding technique.  First it is not possible to apply the technique to
non-linear estimators such as the power spectrum and the bispectrum.  Second,
owing to the structure of the regridding operator, the regridded data are
correlated even if the original data are not.  These correlations are usually
ignored in further processing and the pseudo-data are assumed to be
independent, which results in a poor approximation of the real noise
statistics.  This can be a critical issue with low signal to noise data
\cite{Meimon_et_al-2005-convex_approximation}.

Putting all together, the direct model of the data is affine:
\begin{equation}
  \label{eq:Fourier-plane-data-model}
  \Data = \ModelOp\cdot\VParam + \Error
  %\quad\text{with\ }\ModelOp = \OTFop\cdot\XForm
\end{equation}
with $\Error$ the \emph{error vector}
($\Error[\ell]=\ComplexVis_{j_1,j_2,m}^\ErrorTag$), $\ModelOp =
\OTFop\cdot\XForm$ the linear model operator and 
$\OTFop\in\Complexes^{L\times{}K}$ the OTF operator given by:
\begin{equation}
  \label{eq:otf}
  \OTFop[\ell,k] = 
  \begin{cases}
     \Gain_{j_1,m}^{\star} \, \Gain_{j_2,m}^{}
     & \text{if } \ell\sim(j_1,j_2,m) \in \BaseSet_k\\[1ex]
     0 & \text{else.}
  \end{cases}
\end{equation}

\begin{figure}[!t]
  \centering
  \begin{tabular}{lr}
  \includegraphics[height=33mm]{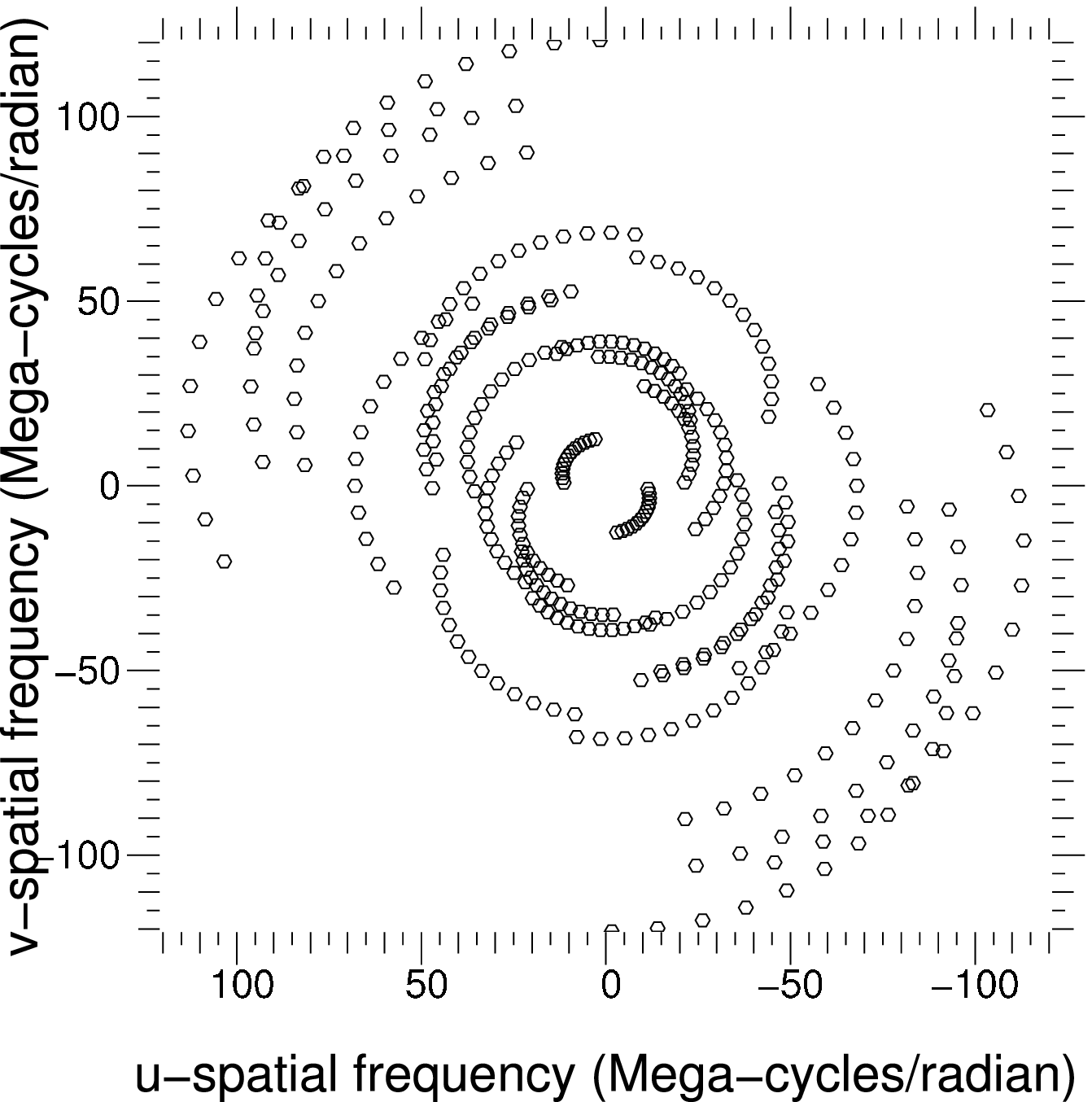} &
  \includegraphics[height=33mm]{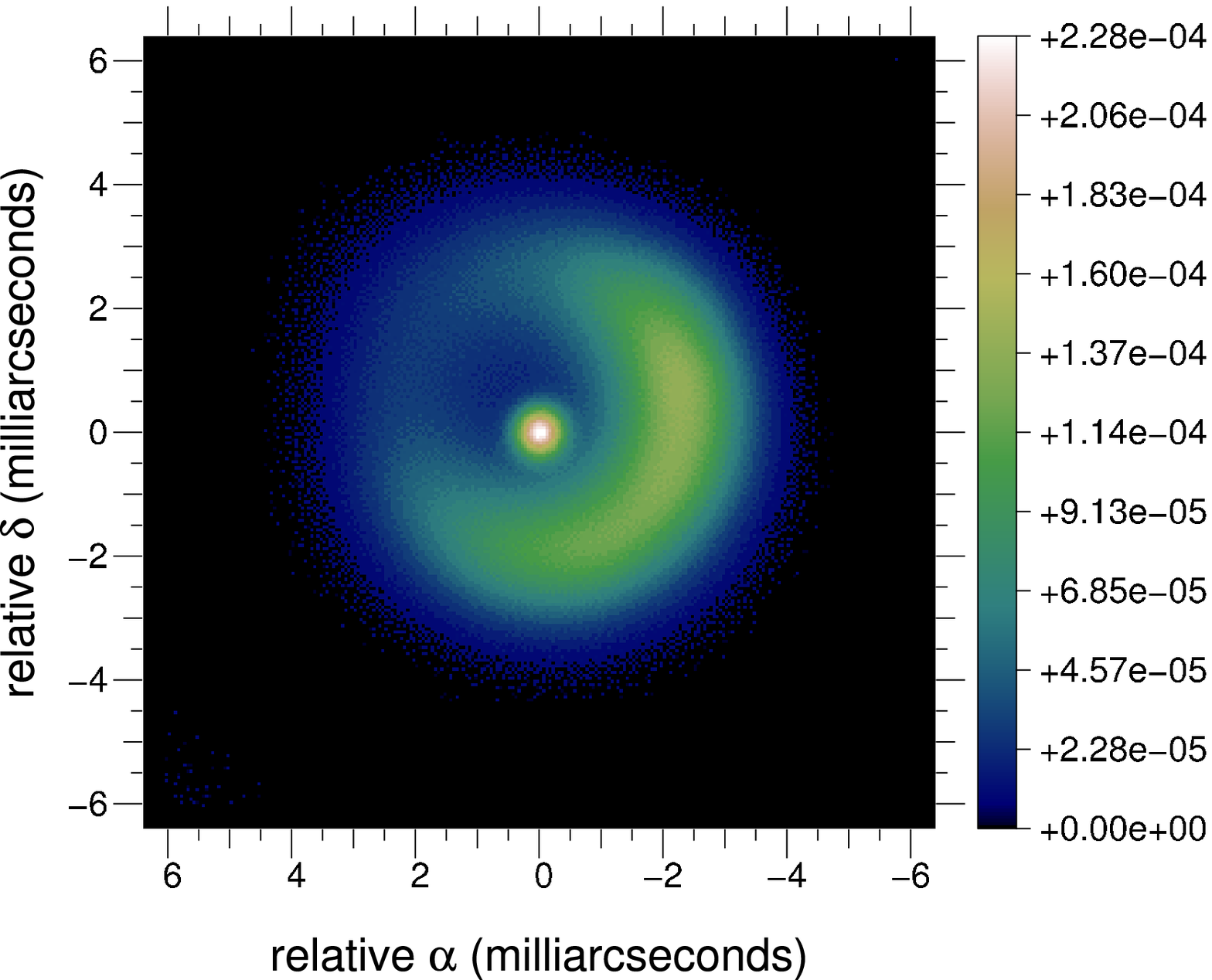} \\[2mm]
  \includegraphics[height=33mm]{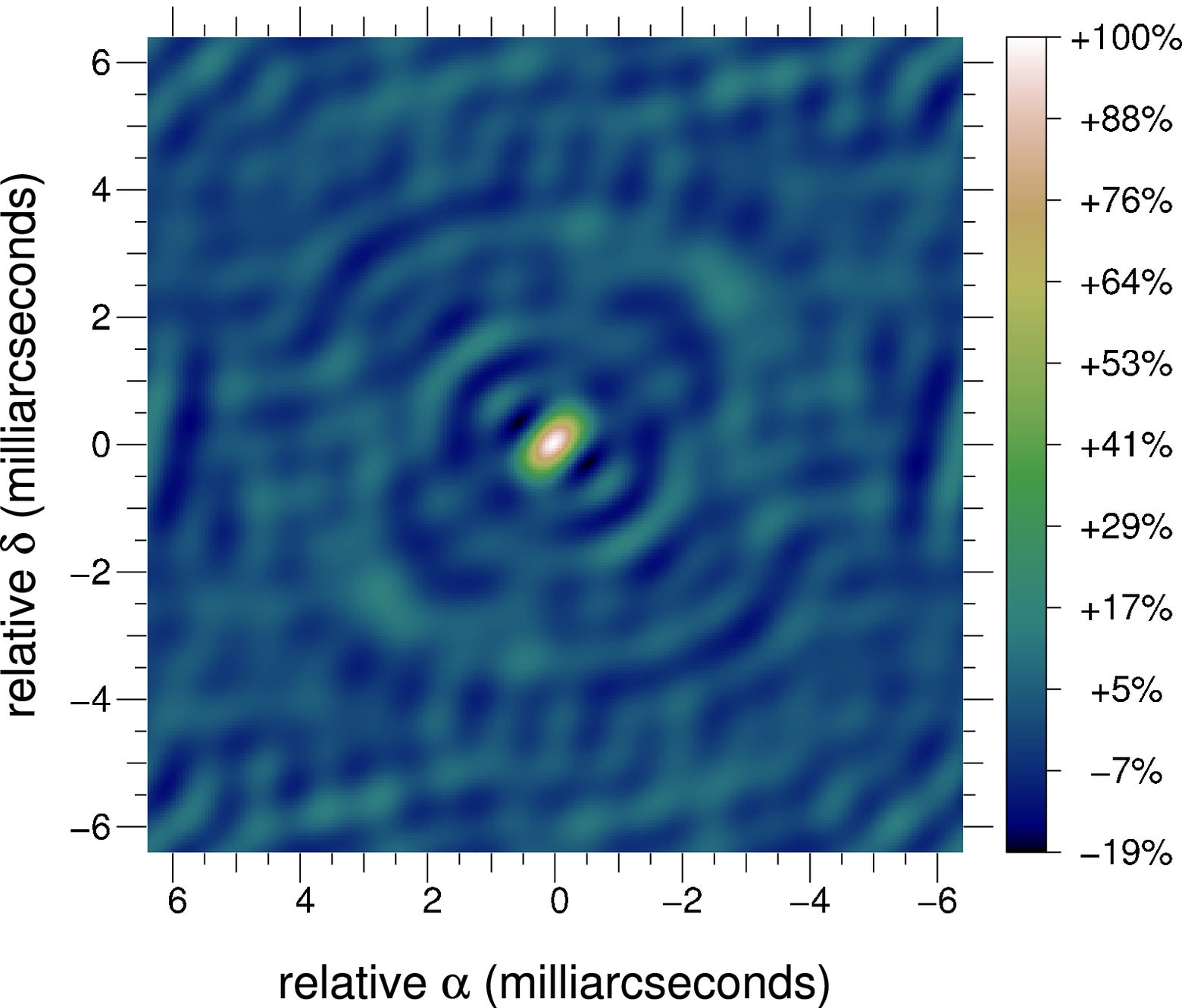} &
  \includegraphics[height=33mm]{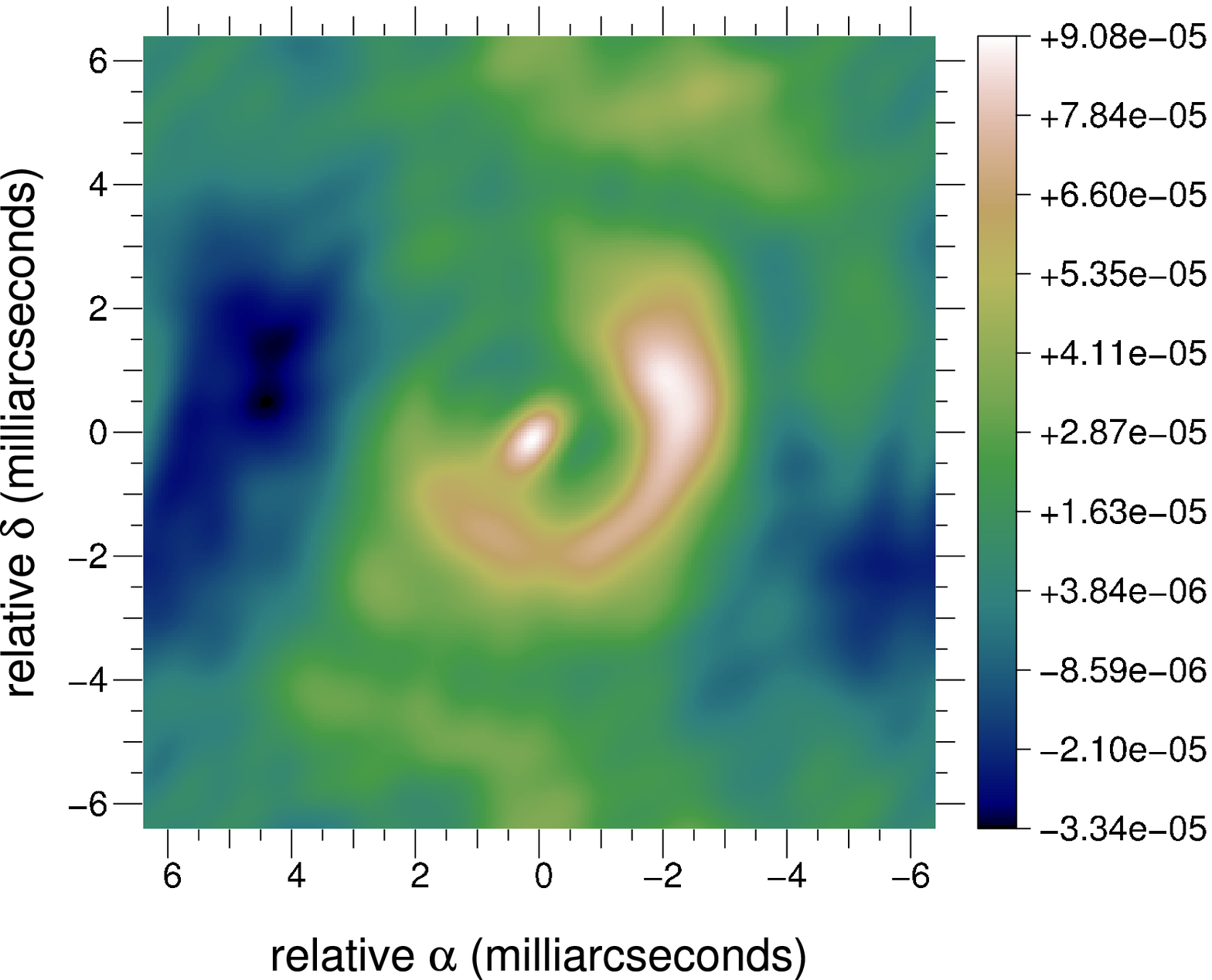}\\
  \end{tabular}
  \caption{Top left: \uv coverage. Top right: observed object.  Bottom left:
    dirty beam.  Bottom right: dirty image.  Object model and \uv coverage are
    from the \emph{2004' Beauty Contest}
    \cite{Lawson_etal-2004-image_beauty_contest}.}
  \label{fig:dirty-map}
\end{figure}

Applying the pseudo-inverse $\XForm^{+} =
\ApodizationOp^{-1}\cdot\DFTop^{-1}\cdot\InterpOp^{+}$ of $\XForm$ to the data
yields the so-called \emph{dirty image} (see \Fig{fig:dirty-map}):
\begin{equation}
  \label{eq:dirty-map}
  \mathring{\Data} = \XForm^{+}\cdot\Data = \M{H}\cdot\VParam
  + \mathring{\Error}
  %VParam_\Tag{dirty} = [\M{M}\T\cdot\Werror\cdot\M{M}]^{\dagger}
  %\cdot\M{M}\T\cdot\Werror\cdot\Data
\end{equation}
where $\mathring{\Error}=\XForm^{+}\cdot\Error$ and
$\M{H}=\XForm^{+}\cdot\OTFop\cdot\XForm$.  Apart from the apodization, $\M{H}$
essentially performs the convolution of the image by the \emph{dirty beam}
(see \Fig{fig:dirty-map}).  From \Eq{eq:Fourier-plane-data-model} and
\Eq{eq:dirty-map}, image reconstruction from interferometric data can be
equivalently seen as a problem of interpolating \emph{missing} Fourier
coefficients or as a problem of deconvolution of the dirty map by the dirty
beam \cite{Giovannelli_Coulais-2005-pos_mix}.

%\begin{figure}[!t]
%  \centering
%  \includegraphics[width=50mm]{alphaBoo-uv-coverage}
%  \caption{\uv coverage at different wavelengths in the H-band for
%    observations of Arcturus in May 2006 with IOTA and achieved with a maximum
%    baseline of 37.7\,m \cite{Lacour_et_al-2008-Arcturus}.}
%  \label{fig:uv-coverage}
%\end{figure}

%==============================================================================
%=========================================================== INVERSE APPROACH =
%==============================================================================

\subsection{Inverse Problem Approach}
\label{sec:inverse-approach}

Since many Fourier frequencies are not measured, fitting the data alone does
not uniquely define the sought image.  Such an ill-posed problem can be solved
by an inverse problem approach \cite{Tarantola-2005-inverse_problem_theory} by
imposing \emph{a priori} constraints to select a unique image among all those
which are consistent with the data.  The requirements for the priors are that
they must help to smoothly interpolate voids in the \uv coverage while
avoiding high frequencies beyond the diffraction limit.  Without loss of
generality, we assume that these constraints are monitored by a penalty
function $\Fprior(\VParam)$ which measures the agreement of the image with the
priors: the lower $\Fprior(\VParam)$, the better the agreement.  In inverse
problem framework, $\Fprior(\VParam)$ is termed as the \emph{regularization}.
Then the parameters $\VParam^{+}$ of the image which best matches the priors
while fitting the data are obtained by solving a constrained optimization
problem:
\begin{equation}
  \label{eq:strict-data}
  \VParam^{+} = \argmin%_{\V{x}\in\FeasibleSet}
  \Fprior(\V{x}) \,,\ \text{subject to:\ }
  \ModelOp\cdot\VParam = \Data \, .
\end{equation}

Other strict constraints may apply.  For instance, assuming the image
brightness distribution must be positive and normalized, the feasible set is:
\begin{equation}
  \label{eq:feasible-set}
  \FeasibleSet =
  \{\VParam\in\Reals^N; \V{\Param} \ge 0, \sum_n \Param_n = 1\}
\end{equation}
where $\V{\Param} \ge 0$ means: $\forall n$, $\Param_n \ge 0$.  Besides, due
to noise and model approximations, there is some expected discrepancy between
the model and actual data.  As for the priors, the distance of the model to
the data can be measured by a penalty function $\Fdata(\VParam)$.  We then
require that, to be consistent with the data, an image must be such that
$\Fdata(\V{x}) \le \DataLevel$ where $\DataLevel$ is set according to the
level of errors:
\begin{equation}
  \label{eq:min-prior-st-data-const}
  \VParam^{+} = \argmin_{\VParam\in\FeasibleSet} \Fprior(\VParam)
  \,,\ \text{subject to:\ }
  \Fdata(\VParam) \le \DataLevel \, .
\end{equation}
The Lagrangian of this constrained optimization problem can be written as:
\begin{equation}
  \mathcal{L}(\VParam; \ell) = \Fprior(\VParam)
  + \ell\,\Bigl(\Fdata(\VParam) - \DataLevel\Bigr)
\end{equation}
where $\ell$ is the Lagrange multiplier associated to the inequality
constraint $\Fdata(\VParam)\le\DataLevel$.  If the constraint is
\emph{active}\footnote{Conversely, the constraint being \emph{inactive} would
  imply that $\ell=0$, which would mean that the data are useless, which is
  hopefully not the case...}, then $\ell>0$ and $\Fdata(\VParam)=\DataLevel$
\cite{Nocedal_Wright-2006-numerical_optimization}.  Dropping the constant
$\DataLevel$ which does not depend on $\VParam$, the solution is obtained by
solving either of the following problems:
\begin{align}
  \VParam^{+}
  &= \argmin_{\VParam\in\FeasibleSet}
     \{\Fprior(\VParam) + \ell\,\Fdata(\VParam)\} \notag \\
% &= \argmin_{\VParam\in\FeasibleSet}
%    \{\Fdata(\VParam) + \mu \, \Fprior(\VParam)\} \notag \\
  &= \argmin_{\VParam\in\FeasibleSet}
     \Fcost(\VParam; \mu) \, , \notag
\end{align}
where 
\begin{equation}
  \label{eq:cost-function}
  \Fcost(\VParam; \mu) = \Fdata(\VParam) + \mu \, \Fprior(\VParam)
\end{equation}
is the penalty function and $\mu=1/\ell>0$ has to be tuned to match the
constraint $\Fdata(\VParam)=\DataLevel$.  Hence we can equivalently consider
that we are solving the problem of maximizing the agreement of the model with
the data subject to the constraint that the priors be below a preset level:
\begin{equation}
  \label{eq:min-data-st-prior-const}
  \VParam^{+} = \argmin_{\VParam\in\FeasibleSet} \Fdata(\VParam)
  \,,\ \text{subject to:\ }
  \Fprior(\VParam) \le \PriorLevel \, .
\end{equation}
For convex penalties and providing that the Lagrange multipliers ($\mu$ and
$\ell$) and the thresholds ($\DataLevel$ and $\PriorLevel$) are set
consistently, the image restoration is achieved by solving either of the
problems in \Eq{eq:min-prior-st-data-const}, \Eq{eq:min-data-st-prior-const}
or by minimizing the penalty function in \Eq{eq:cost-function}.  However,
choosing which of these particular problems to solve can be a deciding issue
for the efficiency of the method.  For instance, if $\Fdata(\VParam)$ and
$\Fprior(\VParam)$ are both \emph{smooth functions}, direct minimization of
$\Fcost(\VParam; \mu)$ in \Eq{eq:cost-function} can be done by using general
purpose optimization algorithms
\cite{Nocedal_Wright-2006-numerical_optimization} but requires to \emph{know}
the value of the Lagrange multiplier.  If the penalty functions are not smooth
or if one wants to have the Lagrange multiplier automatically tuned given
$\DataLevel$ or $\PriorLevel$, specific algorithms must be devised.  As we
will see in the following, specifying the image reconstruction as a
constrained optimization problem provides a very general framework suitable to
describe most existing methods; it however hides important algorithmic details
about the strategy to search the solution.  In the remaining of this section,
we first derive expressions of the data penalty terms and, then, the various
regularizations that have been considered for image reconstruction in
interferometry.

%Going back to the dirty image case (\cf Section~\ref{sec:maximum-likelihood}),
%we see that the degeneracy due to incomplete \uv coverage was solved by using
%the pseudo-inverse which is identical to compute the solution of a
%regularization problem in the limit of a negligible (but still non null) value
%of $\mu$.  The term $\mu\,\norm{\VParam}^2$ is the simplest quadratic
%regularization that could be used, incidentally Tikhonov
%\cite{Tikhonov_Arsenin-1977} was the first to propose this kind of
%regularization to solve inverse problems.

%==============================================================================
%======================================================= DISTANCE TO THE DATA =
%==============================================================================

\subsection{Distance to the Data}
\label{sec:fdata}

The $\ell_2$ norm is a simple means to measure the consistency of the model
image with the data:
\begin{equation}
  \label{eq:fdata-L2}
  \Fdata(\VParam) = \Norm{\Data - \ModelOp\cdot\VParam}^2_2\,.
\end{equation}
However, to account for correlations and for the inhomogeneous quality of the
measurements, the distance to the data has to be defined according to the
statistics of the errors $\Error = \Data - \ModelOp\cdot\VParam$ given the
image model.  Assuming Gaussian statistics, this leads to:
\begin{equation}
  \label{eq:fdata-Gaussian}
  \Fdata(\VParam) =
  \QuadTerm{\Werror}{(\Data - \ModelOp\cdot\VParam)} \,,
\end{equation}
where the weighting matrix $\Werror=\Cerror^{-1}$ is the inverse of the
covariance matrix of the errors.  There is a slight issue because we are
dealing with complex values.  Since complex numbers are just pairs of reals,
complex valued vectors (such as $\Data$, $\Error$ and $\ModelOp\cdot\VParam$)
can be \emph{flattened} into ordinary real vectors (with doubled size) to use
standard linear algebra notation and define the covariance matrix as
$\Cerror=\avg{\Error\cdot\Error\T}$.  This is what is assumed in
\Eq{eq:fdata-Gaussian}.

There are some possible simplifications.  For instance, the complex
visibilities are measured independently, hence the weighting matrix $\Werror$
is block diagonal with $2\times2$ blocks.  Furthermore, if the real and
imaginary parts of a given measured complex visibility are uncorrelated and
have the same variance, then $\Fdata$ takes a simple form:
\begin{equation}
  \label{eq:fdata-Goodman}
  \Fdata(\VParam) = \sum_\ell \Weight_{\ell} \,
  \Abs{\Data[\ell] - (\ModelOp\cdot\VParam)_\ell}^2 \, ,
\end{equation}
where the weights are given by:
\begin{equation}
  \label{eq:fdata-Goodman-weights}
  \Weight_{\ell}
  = \Var\Paren{\Re\Paren{\Data[\ell]}}^{-1}
  = \Var\Paren{\Im\Paren{\Data[\ell]}}^{-1} \, .
\end{equation}
This expression of $ \Fdata(\VParam)$, popularized by Goodman
\cite{Goodman-statistical_optics}, is very commonly used in
radio-interferometry.

\begin{figure}
  \centering
  \includegraphics[height=40mm]{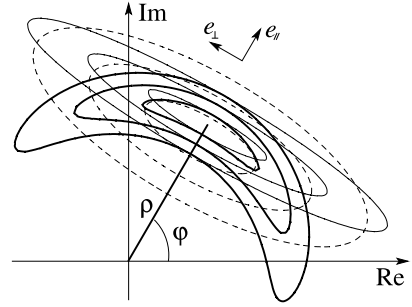}
  \caption{Convex quadratic approximations of complex data. Thick lines:
    isocontours of the log-likelihood $\Fdata$ (at 1, 2 and 3 rms levels) for
    a complex datum with independent amplitude and phase.  Dashed lines:
    isocontours for the global quadratic approximation.  Thin lines:
    isocontours for the local quadratic approximation.}
  \label{fig:quadratic-approximation}
\end{figure}

Real data may however have different statistics.  For instance, the OI-FITS
file exchange format for optical interferometric data assumes that the
amplitude and the phase of complex data (complex visibility or triple product)
are independent \cite{Pauls_et_al-2005-oifits}.  The thick lines in
Fig.~\ref{fig:quadratic-approximation} display the isocontours of the
corresponding log-likelihood which forms a non-convex valley in the complex
plane.  Assuming Goodman statistics would yield circular isocontours in this
figure and is obviously a bad approximation of the true criterion in that
case.  To improve on the Goodman model while avoiding non-convex criteria,
Meimon \etal \cite{Meimon_et_al-2005-convex_approximation} have proposed
quadratic convex approximations of the true log-likelihood (see
Fig.~\ref{fig:quadratic-approximation}) and have shown that their so-called
\emph{local approximation} yields the best results, notably when dealing with
low signal to noise data.  For a complex datum
$\Data[\ell]=\rho_\ell\,\exp(\mathi\,\varphi_\ell)$, their local quadratic
approximation writes:
\begin{equation}
  \label{eq:fdata-local-approx}
  \Fdata(\VParam) = \sum_\ell \Brace{
    \frac{\Re\Paren{e_\ell\,\mathe^{-\mathi\,\varphi_\ell}}^2}
         {\sigma^2_{/\!/,\ell}}
    + \frac{\Im\Paren{e_\ell\,\mathe^{-\mathi\,\varphi_\ell}}^2}
           {\sigma^2_{\perp,\ell}}
  }
\end{equation}
where $\Error = \Data - \ModelOp\cdot\VParam$ denotes the complex residuals
and the variances along and perpendicular to the complex datum vector are:
\begin{align}
  \sigma^2_{/\!/,\ell} &= \Var(\rho_\ell) \\
  \sigma^2_{\perp,\ell} &= \rho_\ell^2\,\Var(\varphi_\ell) \, .
\end{align}
The Goodman model is retrieved when $\rho_\ell^2\,\Var(\varphi_\ell) =
\Var(\rho_\ell)$.

%==============================================================================
%============================================================ MAXIMUM ENTROPY =
%==============================================================================

\subsection{Maximum Entropy Methods}
\label{sec:MEM}

Maximum entropy methods (MEM) are based on the 1950s work of Jaynes on
information theory; the underlying idea is to obtain the least informative
image which is consistent with the data~\cite{Ables-1974-MEM}.  This amounts
to minimizing a criterion like the one in \Eq{eq:cost-function} with
$\Fprior(\VParam) = -S(\VParam)$ where the \emph{entropy} $S(\VParam)$
measures the informational contents of the image $\VParam$.  In this
framework, $\Fprior(\VParam)$ is sometimes called \emph{negentropy}.  Among
all the expressions considered for the negentropy of an image, one of the most
popular is \cite{Gull_Skilling-1984-MEM}:
\begin{equation}
  \label{eq:mem-prior}
  \Fprior(\VParam) = \sum_j \left[
    \Param_j\,\log( \Param_j/\bar{\Param}_j) - \Param_j + \bar{\Param}_j
  \right]
\end{equation}
with $\bar{\VParam}$ the \emph{default image}; that is, the one which would be
recovered in the absence of any data.  Back to information theory, this
expression is similar to the Kullback-Leibler divergence between $\VParam$ and
$\bar{\VParam}$ (with additional terms that cancel for normalized
distributions).  The default image $\bar{\VParam}$ can be taken as being a
flat image, an image previously restored, an image of the same object at a
lower resolution, \etc Narayan \& Nityananda
\cite{Narayan_Nityananda-1986-maximum_entropy_review} reviewed maximum entropy
methods for radio-interferometry imaging and compared the other forms of the
negentropy that have been proposed.  They argued that only non-quadratic
priors can interpolate missing Fourier data and noted that such penalties also
forbid negative pixel values.  The fact that there is no need to explicitly
impose positivity is sometimes put forward by the proponents of these methods.

MEM penalties are usually \emph{separable}, which means that they do not
depend on the ordering of the pixels.  To explicitly enforce some correlation
between close pixels in the sought image $\VParam$ (hence, some
\emph{smoothness}), the prior can be chosen to depend on $\VParam$.  For
instance: $\bar{\VParam}=\M{P}\cdot\VParam$ where $\M{P}$ is some averaging or
smoothing linear operator.  This type of \emph{floating prior} has been used
to loosely enforce constraints such as radial symmetry \cite{Horne1985}.
Alternatively, an intrinsic correlation function (ICF) can be explicitly
introduced by a convolution kernel to impose the correlation structure of the
image~\cite{Gull-1989-MEM}.

% Idea to use a first reconstruction as prior: ?
% Idea to use hidden variables (a map with lower resolution): Gull (1989);
% See also the Pixon methods~\cite{Dixon96,Puetter99} (mais je ne sais plus
% s'il y a du MEM dans le Pixon).

Minimizing the joint criterion in \Eq{eq:cost-function} with entropy
regularization has a number of issues as the problem is highly non-linear and
as the number of unknowns is very large (as many as there are pixels).
Various methods have been proposed, but the most effective algorithm
\cite{Skilling_Bryan-1984-maximum_entropy} seeks for the solution by a
non-linear optimization in a local sub-space of search directions with the
Lagrange multiplier $\mu$ tuned on the fly to match the constraint that
$\Fdata(\VParam)=\DataLevel$.

% There are other algorithms: Cornwell & Evans (1985),

%==============================================================================
%============================================================= Regularization =
%==============================================================================

\subsection{Other Prior Penalties}
\label{sec:regularization}

Bayesian arguments can be invoked to define other types of regularization.
For instance, assuming that the pixels have a Gaussian distribution leads to
quadratic penalties such as:
\begin{equation}
  \label{eq:gauss-prior}
  \Fprior(\VParam) = \QuadTerm{\Cprior^{-1}}{(\VParam - \bar{\VParam})}
\end{equation}
with $\Cprior$ the prior covariance and $\bar{\VParam}$ the prior solution.
Tikhonov's regularization \cite{Tikhonov_Arsenin-1977},
$\Fprior(\VParam)=\norm{\VParam}^2_2$, is the simplest of these penalties.  By
Parseval's theorem, this regularization favors zeroes for unmeasured
frequencies, it is therefore not recommended for image reconstruction in
interferometry.  Yet, this does not rule out all quadratic priors. For
instance, \emph{compactness} is achieved by a very simple quadratic penalty:
\begin{equation}
  \label{eq:fov-penalty}
  \Fprior(\VParam) = \sum_{n} w^\PriorTag_n\,\Param_n^2 \, ,  
\end{equation}
where the weights are increasing with the distance to the center of the image
thus favoring structures concentrated within this part of the image.  Under
strict normalization constraint and in the absence of any data, the default
image given by this prior is $\bar{x}_n\propto1/w^\PriorTag_n$ where the
factor comes from the normalization requirement
\cite{LeBesnerais_et_al-2008-interferometry}.  Although simple, this
regularizer, coupled with the positivity constraint, can be very effective as
shown by Fig.~\ref{fig:regularization-types}b.  Indeed, smooth Fourier
interpolation follows from the compactness of the brightness distribution
which is imposed by $\Fprior(\VParam)$ in \Eq{eq:fov-penalty} and by the
positivity as it plays the role of a floating support.

%Hence
%\begin{equation}
%  \label{eq:fov-weight}
%  w_\PriorTag(\VDirn) = 1 + \frac{\Dirn_1^2 + \Dirn_2^2}{4\,\Gamma^2}  
%\end{equation}
%yields a default image with radial symmetry and Lorentzian shape with full
%width at half minimum (FWHM) equals to $\Gamma$. 

Other prior penalties commonly used in image restoration methods can be useful
for interferometry.  For instance, \emph{edge-preserving smoothness} is
achieved by:
\begin{equation}
  \label{eq:edge-preserving-roughness}
  \Fprior(\VParam) = \sum_{n_1,n_2}
  \sqrt{\epsilon^2 + \abs{\nabla\VParam}_{n_1,n_2}^2}
\end{equation}
where $\epsilon>0$ is a chosen threshold and $\abs{\nabla\VParam}^2$ is the
squared magnitude of the spatial gradient of the image:
\begin{displaymath}
  \label{eq:spatial-gradient-magnitude}
  \abs{\nabla\VParam}_{n_1,n_2}^2 =
  (x_{n_1+1,n_2} - x_{n_1,n_2})^2 + (x_{n_1,n_2+1} - x_{n_1,n_2})^2 \, .
\end{displaymath}
The penalization in \Eq{eq:edge-preserving-roughness} behaves as a quadratic
(resp.\ linear) function where the magnitude of the spatial gradient is small
(resp.\ large) compared to $\epsilon$.  Thus reduction of small local
variations without penalizing too much strong sharp features is achieved by
this regularization.
% Other specific expressions for the edge-preserving penalty have been
% considered (\eg, \cite{Mugnier_Fusco_etal-2004-JOSAA-Mistral}).
In the limit $\epsilon\rightarrow0$, edge-preserving smoothness behaves like
\emph{total variation} \cite{Strong_Chan-2003-total_variation} which has
proved successful in imposing sparsity.

%==============================================================================
%====================================================================== CLEAN =
%==============================================================================

\subsection{\textsc{Clean} method}
\label{sec:Clean}

% For the proponents of maximum entropy methods, only non-quadratic
% regularizations can lever the degeneracies in image reconstruction problem;
% that is, interpolate the complex visibility for missing spatial frequencies
% \cite{Narayan_Nityananda-1986-maximum_entropy_review}.  In fact, the
% necessary condition for an effective regularization in this context is not
% so much to avoid noise amplification -- though some kind of smoothness
% constraint may be necessary to suppress high frequencies beyond the \uv
% coverage -- but mostly to set constraints so that Fourier interpolation of
% missing frequencies is done smoothly.

% hence = donc
% thus = ainsi

Favoring images with a limited number of significant pixels is a way to avoid
the degeneracies of image reconstruction from sparse Fourier coefficients.
This could be formally done by searching for the least $\ell_0$-norm image
consistent with the data; hence using $\Fprior(\VParam)=\norm{\VParam}_0$.
However, due to the number of parameters, minimizing the resulting mixed
criterion is a combinatorial problem which is too difficult to solve directly.
\Clean algorithm \cite{Fomalont73, Hogbom-1974-CLEAN} implements a
\emph{matching pursuit} strategy to attempt to find this kind of solution.
The method proceeds iteratively as follows. Given the data in the form of a
dirty image, the location of the brightest point source that best explains the
data is searched.  The model image is then updated by a fraction of the
intensity of this component, and this fraction times the dirty beam is
subtracted from the dirty image.  The procedure is repeated for the new
residual dirty image which is searched for evidence of another point-like
source.  When the level of the residuals becomes smaller than a given
threshold set from the noise level, the image is convolved with the
\emph{clean beam} (usually a Gaussian shaped PSF) to set the resolution
according to the extension of the \uv coverage.  Once most point sources have
been removed, the residual dirty image is essentially due to the remaining
extended sources which may be smooth enough to be insensitive to the
convolution by the dirty beam.  Hence adding the residual dirty image to the
clean image produces a final image consisting in compact sources (convolved by
the clean beam) plus smooth extended components.  Although designed for point
sources, \Clean works rather well for extended sources and remains one of the
preferred methods in radio-interferometry.

% While Clean is clearly designed for point-like sources, it works
% to some extend on extended emission... 

It has been demonstrated that the matching pursuit part of \Clean is
equivalent to an iterative deconvolution with early stopping
\cite{Schwarz:1978:CLEAN} and that it is an approximate algorithm for
obtaining the image of minimum total flux consistent with the observations
\cite{Marsh_Richardson-1987-CLEAN_objective_function}.  Hence, under the
non-negativity constraint, this would, at best, yield the least $\ell_1$-norm
image consistent with the data.  This objective is supported by recent results
in \emph{Compressive Sensing}
\cite{Candes_Romberg_Tao-2006-robust_uncertainty} showing that, in most
practical cases, regularization by the $\ell_1$-norm of $\VParam$ enforces the
sparsity of the solution.  However, the matching pursuit strategy implemented
by \Clean is slow, it has some instabilities and it is known to be sub-optimal
\cite{Marsh_Richardson-1987-CLEAN_objective_function}.

% Note that using an $\ell_1$-norm regularization has no effects when
% combined with positivity and normalization constraints, see
% \Eq{eq:feasible-set}.

% The positivity constraint is also a means to enforce a floating support
% and has proved its effectiveness for astronomical images.

% The algorithm of Giovannelli \& Coulais
% \cite{Giovannelli_Coulais-2005-pos_mix} (described in next section) can be
% seen as an improved \Clean method which is explicitly based on the
% minimization of a convex criterion.

\subsection{Other methods}
\label{sec:misc-image-reconstruction-methods}

This section briefly reviews other image reconstruction methods applied in
astronomical interferometry.

% Most of these methods are not specific to interferometry, they were also
% used for general deconvolution.

\subsubsection{Multi-resolution}

These methods aim at reconstructing images with different scales.  They
basically rely on recursive decomposition of the image in low and high
frequencies.
% Several authors improved over \Clean by using a multi-resolution approach.
\emph{Multi-resolution Clean} \cite{Wakker_Schwarz-1988-multiresolution_Clean}
first reconstructs an image of the broad emission and then iteratively updates
this map at full resolution as in the original \Clean algorithm.  This
approach has been generalized by using a wavelet expansion to describe the
image --- which could be formally expressed in terms of
\Eq{eq:general-image-model} --- and achieved multi-resolution deconvolution by
a matching pursuit algorithm applied to the wavelet coefficients and such that
the solution satisfies positivity and support constraints
\cite{Starck_et_al-1994-aperture_synthesis}.  The \emph{multi-scale Clean}
algorithm \cite{Cornwell-2008-multiscale_CLEAN} explicitly describes the image
as a sum of components with different scales and makes use of a weighted
matching pursuit algorithm to search for the scale and position of each image
update.  The main advantages of \emph{multi-scale Clean} are its ability to
leave very few structures in the final residuals and to correctly estimate the
total flux of the observed object.  This method is widely used in
radio-astronomy and is part of standard data processing packages
\cite{Rich_et_al-2008-multiscale_CLEAN_comparison}.  In the context of MEM,
the \emph{multi-channel maximum entropy} image reconstruction
method~\cite{Weir-1992-multichannel_MEM} introduces a multi-scale structure in
the image by means of different intrinsic correlation
functions~\cite{Gull-1989-MEM}.  The reconstructed image is then the sum of
several extended sources with different levels of correlation.  This approach
was extended by using a pyramidal image decomposition
\cite{Bontekoe_at_al-1994-pyramid} or wavelet expansions
\cite{Pantin_Starck-1996-multiscale_maxent,
  Starck_et_al-2001-multiresolution}.

\subsubsection{Wipe method}

The \Wipe method \cite{Lannes_et_al-1997-Clean_and_Wipe} is a regularized fit
of the interferometric data under positivity and support constraints.  The
model image is given by \Eq{eq:general-image-model} using an equally-spaced
grid and the effective resolution is explicitly set by the basis function
$\BasisFunc(\VDirn)$, the so-called \emph{neat beam}, with an additional
penalty to avoid super resolution.  The image parameters are the ones that
minimize:
\begin{displaymath}
  \label{eq:Wipe-criterion}
  f_\Tag{Wipe}(\VParam)
  = \sum_{\ell} w_\ell \, \abs{\hat{\BasisFunc}_\ell\,\Data[\ell]
    - (\ModelOp\cdot\VParam)_\ell}^2
  + \hspace{-.3em}\sum_{k, \abs{\Freq_k} > \Freq_\Tag{eff}}\hspace{-.3em}
  \abs{(\DFTop\cdot\VParam)_k}^2
\end{displaymath}
with $\Data$ the calibrated complex visibility data, $\hat{\BasisFunc}_\ell$
the Fourier transform of the neat beam at the spatial frequency of datum
$\Data[\ell]$, $\Freq_\Tag{eff} \gtrsim \mathrm{sup}_{\Freq \in
  \FreqSet}\abs{\Freq}$ an effective cutoff frequency, $\DFTop$ the Fourier
transform operator, $\ModelOp$ the model matrix given by
\Eq{eq:Fourier-plane-data-model} accounting for the sub-sampled Fourier
transform and the neat beam, and $w_\ell = 1/(\sigma^2_\ell\,\sum_{\ell'}
\sigma^{-2}_{\ell'})$ where $\sigma^2_\ell =
\abs{\hat{\BasisFunc}_\ell}^2\,\Var(\Data[\ell])$ assuming the Goodman
approximation.  In the criterion minimized by \Wipe, one can identify the
distance of the model to the data and a regularization term.  There is no
hyper-parameter to tune the level of this latter term.  The optimization is
done by a conjugate gradient search with a stopping criterion derived from the
analysis of the conditioning of the problem.  This analysis is built up during
the iterations.

\subsubsection{Bi-model method}

The case of an image model explicitly mixing extended source and point sources
has also been addressed \cite{Magain_et_al-1998-correct_sampling,
  Pirzkal_et_al-2000-photometric_restoration} and more recently
\cite{Giovannelli_Coulais-2005-pos_mix}.  The latter have considered an image
$\VParam = \VParam_\Tag{e} + \VParam_\Tag{p}$ made of two maps:
$\VParam_\Tag{e}$ for \emph{extended} structures and $\VParam_\Tag{p}$ for
\emph{point-like} components.  The maps $\VParam_\Tag{e}$ and
$\VParam_\Tag{p}$ are respectively regularized by imposing smoothness and
sparsity.  With additional positivity and, optionally, support constraints, it
turns out that the two kinds of regularization can be implemented by quadratic
penalties. Their method amounts to minimize:
%\begin{align}
%  \label{eq:Giovannelli_Coulais-penalty}
%  f_\Tag{mix}(\VParam_\Tag{e},\VParam_\Tag{p})
%  &= \Norm{\V{y} - \DFTop\cdot(\VParam_\Tag{e} + \VParam_\Tag{p})}^2
%  + \lambda_\Tag{s}\,\sum_n [\VParam_\Tag{p}]_n
%  + \epsilon_\Tag{s}\,\sum_n [\VParam_\Tag{p}]^2_n \notag \\
%  &\quad +  \lambda_\Tag{c}\,\sum_n
%  \Bigl([\VParam_\Tag{e}]_{n+1} - [\VParam_\Tag{e}]_n\Bigr)^2
%  + \epsilon_\Tag{m}\,\Bigl(\sum_n [\VParam_\Tag{e}]_n\Bigr)^2
%\end{equation}
\begin{align}
  %\label{eq:Giovannelli_Coulais-penalty}
  f_\Tag{mix}(\VParam_\Tag{e},\VParam_\Tag{p})
  &= \Norm{\Data - \ModelOp\cdot(\VParam_\Tag{e} + \VParam_\Tag{p})}^2
  + \lambda_\Tag{s}\,\V{c}\T\cdot\VParam_\Tag{p} \notag \\
  & \hspace{1em} + \epsilon_\Tag{s}\,\norm{\VParam_\Tag{p}}^2
  +  \lambda_\Tag{c}\,\norm{\VParam_\Tag{e}}_{\Cprior}^2
  + \epsilon_\Tag{m}\,(\V{c}\T\cdot\VParam_\Tag{e})^2 \, \notag
\end{align}
with $\norm{\VParam_\Tag{e}}^2_{\Cprior}$ a local finite difference norm
similar to \Eq{eq:gauss-prior}, and $\V{c}$ a vector with all components set
to one, hence: $\V{c}\T\cdot\VParam = \sum_n \Param_n$.  There are 4 tuning
parameters for the regularization terms: $\lambda_\Tag{s}\ge0$ and
$\epsilon_\Tag{s}>0$ control the sparsity of $\VParam_\Tag{p}$,
$\lambda_\Tag{c}>0$ controls the level of smoothness in the extended map
$\VParam_\Tag{e}$, and $\epsilon_\Tag{m}>0$ (or $\epsilon_\Tag{m}\ge0$ if
there is a support constraint) insures strict convexity of the regularization
\wrt $\VParam_\Tag{e}$.  Circulant approximations are used to implement a very
fast minimization of $f_\Tag{mix}(\VParam_\Tag{e},\VParam_\Tag{p})$ under the
constraints that $\VParam_\Tag{e}\ge0$ and $\VParam_\Tag{p}\ge0$
\cite{Giovannelli_Coulais-2005-pos_mix}.

%==============================================================================
%=========================================================== SELF-CALIBRATION =
%==============================================================================

\subsection{Self-Calibration}
\label{sec:self-calibration}

When the OTF cannot be calibrated (\eg there is no reference or the OTF is
significantly varying due to the turbulence), the problem is not only to
derive the image parameters $\VParam$, but also the unknown complex
throughputs $\VGain$.  As there is no correlation between the throughputs and
the observed object, the inverse approach leads to solve:
\begin{align}
  \label{eq:blind-map-problem}
  (\VParam,\VGain)^\BestTag
  &= \argmin_{\VParam\in\FeasibleSet,\VGain} \left\{\Fdata(\VParam,\VGain)
  + \mu_\Tag{img}\,\Fprior^\Tag{img}(\VParam)\right.\notag\\
  &\hspace{8.7em}\quad \left.
  {+}\:\mu_\Tag{gain}\,\Fprior^\Tag{gain}(\VGain) \right\}
\end{align}
with $\mu_\Tag{img}\,\Fprior^\Tag{img}(\VParam)$ and
$\mu_\Tag{gain}\,\Fprior^\Tag{gain}(\VGain)$ the regularization terms for the
image parameters and for the complex throughputs.  The latter can be derived
from prior statistics about the turbulence \cite{Roddier1981}.  In principle,
\emph{global} optimization should be required to minimize the non-convex
criterion in \Eq{eq:blind-map-problem}.  Fortunately, a simpler strategy based
on alternate minimization with respect to $\VParam$ only and then with respect
to $\VGain$ only has proved effective to solve this problem.  This method has
been called \emph{self-calibration} because it uses the current estimate of
the sought image as a reference source to calibrate the throughputs.  The
algorithm begins with an initial image $\VParam^{[0]}$ and repeats until
convergence the following steps (starting with $n=1$ and incrementing $n$
after each iteration):
\begin{enumerate}
\item \emph{Self-calibration step.} Given the image $\VParam^{[n-1]}$, find
  the \emph{best} complex throughputs $\VGain^{[n]}$ by solving:
  \begin{displaymath}
    \VGain^{[n]} = \argmin_{\VGain} \left\{
    \Fdata\bigl(\VParam^{[n-1]}, \VGain\bigr)
    + \mu_\Tag{gain}\,\Fprior^\Tag{gain}(\VGain) \right\} \, .
  \end{displaymath}
\item \emph{Image reconstruction step.} Apply image reconstruction algorithm
  to recover a new image estimate given the data and the complex throughputs:
  \begin{displaymath}
    \VParam^{[n]} = \argmin_{\VParam} \left\{
    \Fdata\bigl(\VParam, \VGain^{[n]}\bigr)
    + \mu_\Tag{img}\,\Fprior^\Tag{img}(\VParam) \right\} \, .
  \end{displaymath}
\end{enumerate}  
Note that any image reconstruction algorithm described previously can be used
in the second step of the method.  The criterion in \Eq{eq:blind-map-problem}
being non-convex, the solution should depend on the initialization.  Yet, this
does not appear to be an issue in practice even if simple local optimization
methods are used to solve the self-calibration step (such as the one recently
proposed by Lacour \etal \cite{Lacour_etal-2007-pupil_remapping_imaging}).

Self-calibration was initially proposed by Readhead \& Wilkinson
\cite{Readhead_Wilkinson-1978-VLBI} to derive missing Fourier phase
information from phase closure data, and the technique was later improved by
Cotton \cite{Cotton-1979-VLBI}.  Schwab \cite{Schwab-1980-self_calibration}
was the first to solve the problem by explicitly minimizing a non-linear
criterion similar to $\Fdata(\VParam,\VGain)$ in \Eq{eq:fdata-Goodman}.
Schwab's approach was further improved by Cornwell \& Wilkinson
\cite{Cornwell_Wilkinson-1981-self_calibration} who introduced priors for the
complex gains, that is, the term $\mu_\Tag{gain}\,\Fprior^\Tag{gain}(\VGain)$
in the global penalty.  However, for most authors, no priors about the
throughputs are assumed, hence $\mu_\Tag{gain}=0$.

Self-calibration is a particular case of the \emph{blind}, or \emph{myopic},
deconvolution methods \cite{Campisi_Egiazarian-2007-blind_deconvolution} that
have been developed to improve the quality of blurred images when the point
spread function (PSF) is unknown.  Indeed, when the PSF can be completely
described by phase aberrations in the pupil plane, blind deconvolution amounts
to solving the same problem as self-calibration
\cite{Schulz-1993-multiframe_blind_deconvolution}.

\begin{figure}[!t]
  \centering
  \begin{tabular}{lr}
  \includegraphics[height=33mm]{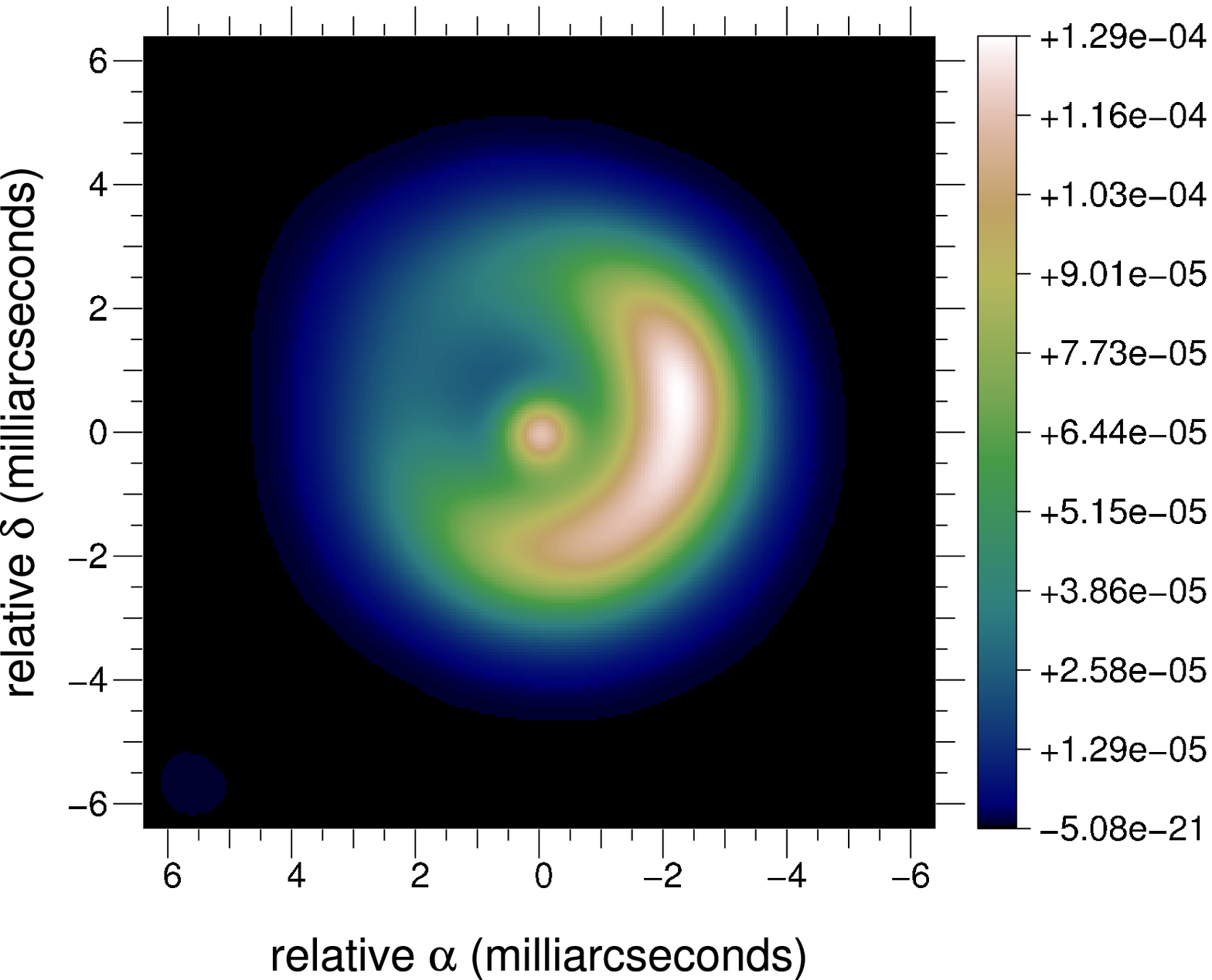} &
  \includegraphics[height=33mm]{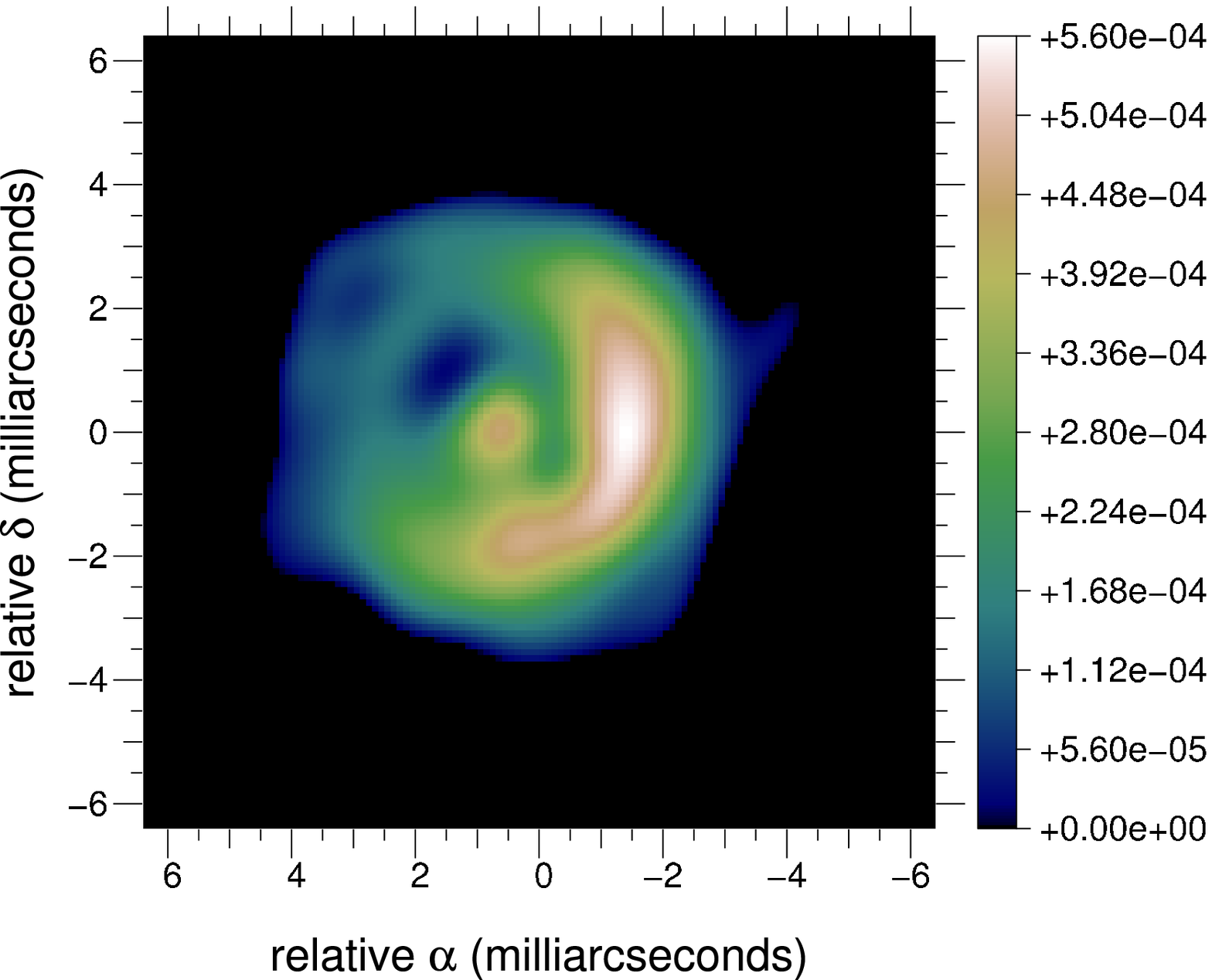} \\[2mm]
  \includegraphics[height=33mm]{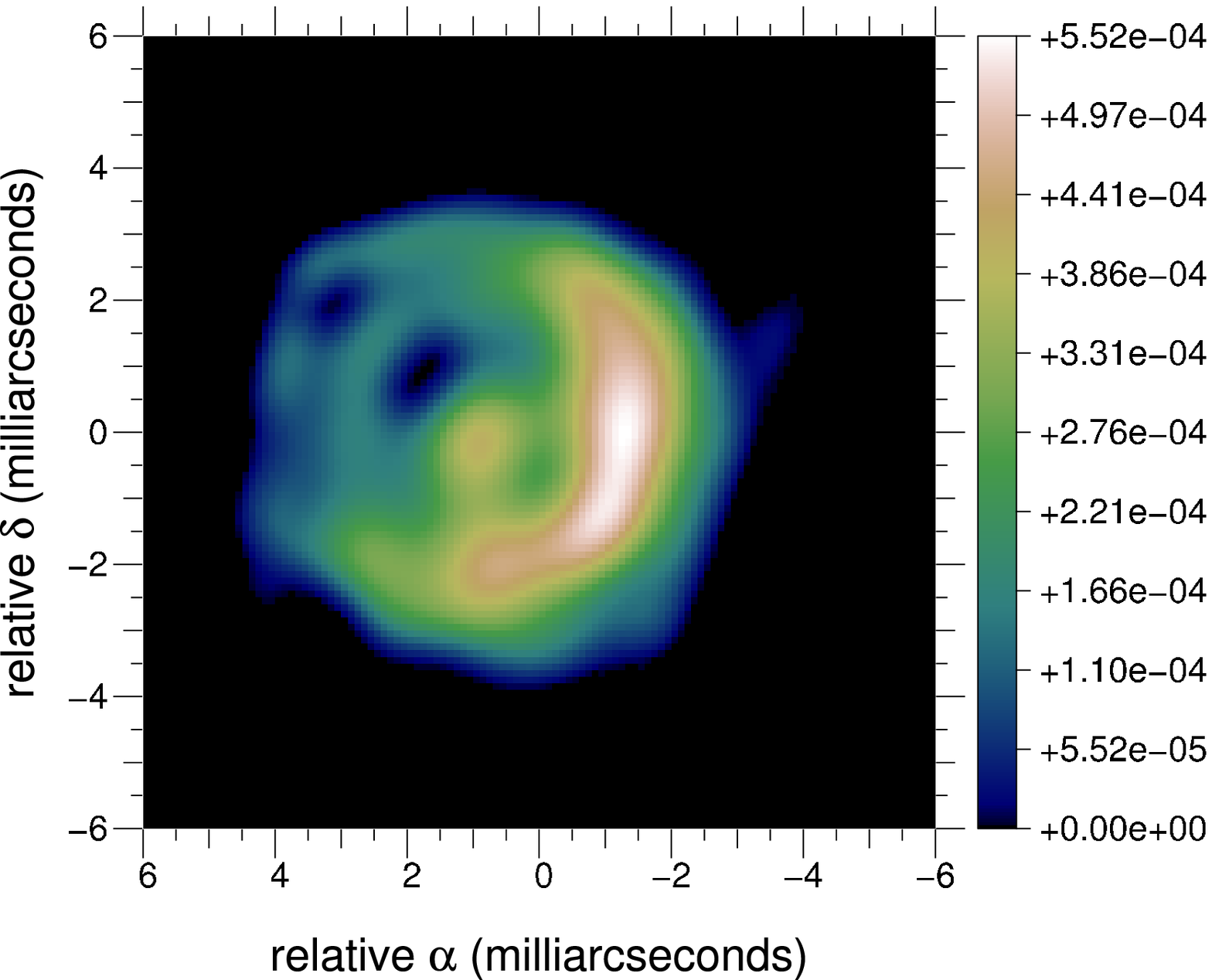} &
  \includegraphics[height=33mm]{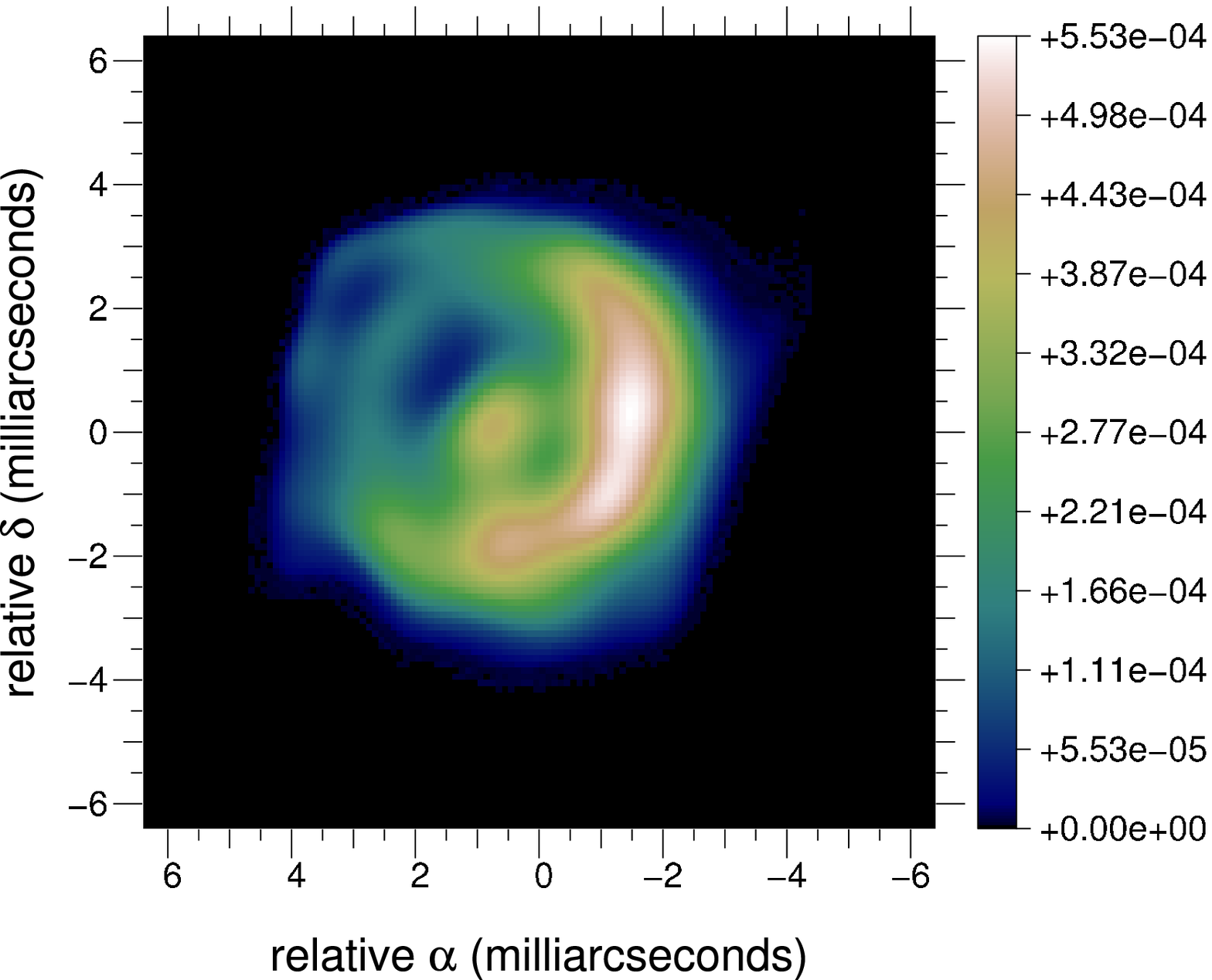}\\
  \end{tabular}
  \caption{Image reconstruction with various types of regularization.  From
    top-left to bottom-right: (a) original object smoothed to the resolution
    of the interferometer ($\mathrm{FWHM}\sim15\,\mathrm{mas}$); (b)
    reconstruction with a quadratic regularization given by
    Eq.~(\protect\ref{eq:fov-penalty}) and which imposes a compact field of
    view; (c) reconstruction with edge-preserving regularization as in
    Eq.~(\protect\ref{eq:edge-preserving-roughness}); (d) reconstruction with
    maximum entropy regularization as in Eq.~(\protect\ref{eq:mem-prior}).
    All reconstructions by algorithm \Mira and from the power spectrum and the
    phase closures.}
  \label{fig:regularization-types}
\end{figure}

%%%%%%%%%%%%%%%%%%%%%%%%%%%%%%%%%%%%%%%%%%%%%%%%%%%%%%%%%%%%%%%%%%%%%%%%%%%%%%%
%%%%%%%%%%%%%%%%%%%%%%%%%%%%%%%%%%%%%%%%%%%%%%%%%%%%%% OPTICAL INTERFEROMETRY %
%%%%%%%%%%%%%%%%%%%%%%%%%%%%%%%%%%%%%%%%%%%%%%%%%%%%%%%%%%%%%%%%%%%%%%%%%%%%%%%

\section{Image Reconstruction from Non-Linear Data}
\label{sec:optical-interferometry}

At optical wavelengths, the complex visibilities (whether they are calibrated
or not) are not directly measurable, the available data (\cf Section
\ref{sec:interferometric-data}) are the power spectrum, the bispectrum and/or
the phase closure.  Image reconstruction algorithms can be designed following
the same inverse problem approach as previously.  In particular, the
regularization can be implemented by the same $\Fprior$ penalties as in
Section~\ref{sec:image-reconstruction}.  However, the direct model of the data
is now non-linear and specific expressions to implement $\Fdata$ have to be
derived.  The non-linearity has also some incidence on the optimization
strategy.

\subsection{Data Penalty}

The power spectrum, the bispectrum, and the phase closure data have
non-Gaussian statistics: the power spectrum is a positive quantity, the phase
closure is wrapped in $(-\pi,+\pi]$, \etc Most algorithms however make use of
quadratic penalties \wrt the measurements which implies Gaussian statistics in
a Bayesian framework.  Another assumption generally made is the independence
of the measurements, which leads to \emph{separable} penalties.  Under such
approximations, the penalty \wrt the power spectrum data writes:
\begin{equation}
  \label{eq:powerspectrum-penalty}
  \Fdata^\PowerspectrumTag(\VParam) = \sum_{m, j_1 < j_2}
  \frac{  \Paren{\Powerspectrum_{j_1,j_2,m}^\DataTag
   - \Powerspectrum_{j_1,j_2,m}^\ModelTag(\VParam)}^2
  }{\Var\Paren{\Powerspectrum_{j_1,j_2,m}^\DataTag}} \, ,
\end{equation}
with $\Powerspectrum_{j_1,j_2,m}^\ModelTag(\VParam) =
\abs{\FT{\Image}(\VFreq_{j_1,j_2,m})}^2$ the model of the power spectrum.  For
the penalty \wrt the bispectrum data, there is the additional difficulty to
deal with complex data.  The Goodman approximation
\cite{Goodman-statistical_optics} yields:
\begin{equation}
  \label{eq:bispectrum-penalty}
  \Fdata^\BispectrumTag(\VParam) =
  \hspace{-1em}\sum_{m, j_1 < j_2 < j_3}\hspace{-1em}
  w^\BispectrumTag_{j_1,j_2,j_3,m}\,
  \Abs{\Bispectrum_{j_1,j_2,j_3,m}^\DataTag
   - \Bispectrum_{j_1,j_2,j_3,m}^\ModelTag(\VParam)}^2
\end{equation}
with $\Bispectrum_{j_1,j_2,j_3,m}^\ModelTag(\VParam) =
\FT{\Image}(\VFreq_{j_1,j_2,m}) \, \FT{\Image}(\VFreq_{j_2,j_3,m}) \,
\FT{\Image}(\VFreq_{j_3,j_1,m})$ the model of the bispectrum and weights
derived from the variance of the bispectrum data.  An expression similar to
that in \Eq{eq:fdata-local-approx} can be derived for bispectrum data with
independent modulus and phase errors.  To account for phase wrapping, Haniff
\cite{Haniff1991} proposed to define the penalty \wrt the phase closure data
as:
\begin{equation}
  \label{eq:Haniff-phase-closure-penalty}
  \Fdata^\PhaseClosureTag(\VParam) =
  \hspace{-1em}\sum_{m, j_1 < j_2 < j_3}\hspace{-1em}
  \frac{
    \arc^2\Paren{\PhaseClosure_{j_1,j_2,j_3,m}^\DataTag
   - \PhaseClosure_{j_1,j_2,j_3,m}^\ModelTag(\VParam)}
  }{\Var\Paren{\PhaseClosure_{j_1,j_2,j_3,m}^\DataTag}}
\end{equation}
with $\PhaseClosure_{j_1,j_2,j_3,m}^\ModelTag(\VParam) =
\VisPhase(\VFreq_{j_1,j_2,m}) + \VisPhase(\VFreq_{j_2,j_3,m}) +
\VisPhase(\VFreq_{j_3,j_1,m})$ the model of the phase closure.  This penalty
is however not continuously differentiable \wrt $\VParam$, which can prevent
the convergence of optimization algorithms.  This problem can be avoided by
using the complex phasors \cite{Thiebaut-2008-Marseille}:
\begin{equation}
  \label{eq:Mira-phase-closure-penalty}
  \Fdata^\PhaseClosureTag(\VParam) =
  \hspace{-1em}\sum_{m, j_1 < j_2 < j_3}\hspace{-1em}
  \frac{
    \bigl\vert
      \mathe^{\mathi\,\PhaseClosure_{j_1,j_2,j_3,m}^\DataTag}
    - \mathe^{\mathi\,\PhaseClosure_{j_1,j_2,j_3,m}^\ModelTag(\VParam)}
    \bigr\vert^2
  }{\Var\Paren{\PhaseClosure_{j_1,j_2,j_3,m}^\DataTag}} \, ,
\end{equation}
which is approximately equal to the penalty in
\Eq{eq:Haniff-phase-closure-penalty} in the limit of small phase closure
errors.

Depending on which set of data is available, and assuming that the different
types of data have statistically independent errors, the \emph{total} penalty
\wrt the data is simply a sum of some of the penalties given by
equations~(\ref{eq:powerspectrum-penalty})--(\ref{eq:Mira-phase-closure-penalty}).
For instance, to fit the power spectrum and the phase closure data:
\begin{equation}
  \Fdata(\VParam) = \Fdata^\PowerspectrumTag(\VParam) +
  \Fdata^\PhaseClosureTag(\VParam) \, .
\end{equation}
%Note that existing algorithms can impose specific expressions for
%$\Fdata(\VParam)$ and $\Fprior(\VParam)$.

\subsection{Image Reconstruction Algorithms}
\label{sec:new-methods}

We describe here the image reconstruction methods used with some success on
realistic optical interferometric data in astronomy and which can be
considered as ready to process real data.  In addition to cope with sparse
Fourier data, these methods were specifically designed to tackle the
non-linear direct model, to account for the particular statistics of the data
\cite{Meimon_et_al-2005-convex_approximation} and to handle the new data
format \cite{Pauls_et_al-2005-oifits}.  These image reconstruction methods can
all be formally described in terms of a criterion to optimize, perhaps under
some strict constraints, and an optimization strategy.  Some of these
algorithms have clearly inherited from methods previously developed: BSMEM
\cite{Buscher-1994-BSMEM}, the \emph{Building-Block Method}
\cite{Hofmann_Weigelt-1993-building_blocks} and \Wisard
\cite{Meimon_et_al-2005-weak_phase_imaging} are respectively related to MEM
(\cf Section \ref{sec:MEM}), \Clean (\cf Section \ref{sec:Clean}) and
self-calibration (\cf Section \ref{sec:self-calibration}).

\begin{figure}[!t]
  \centering
  \includegraphics[height=33mm]{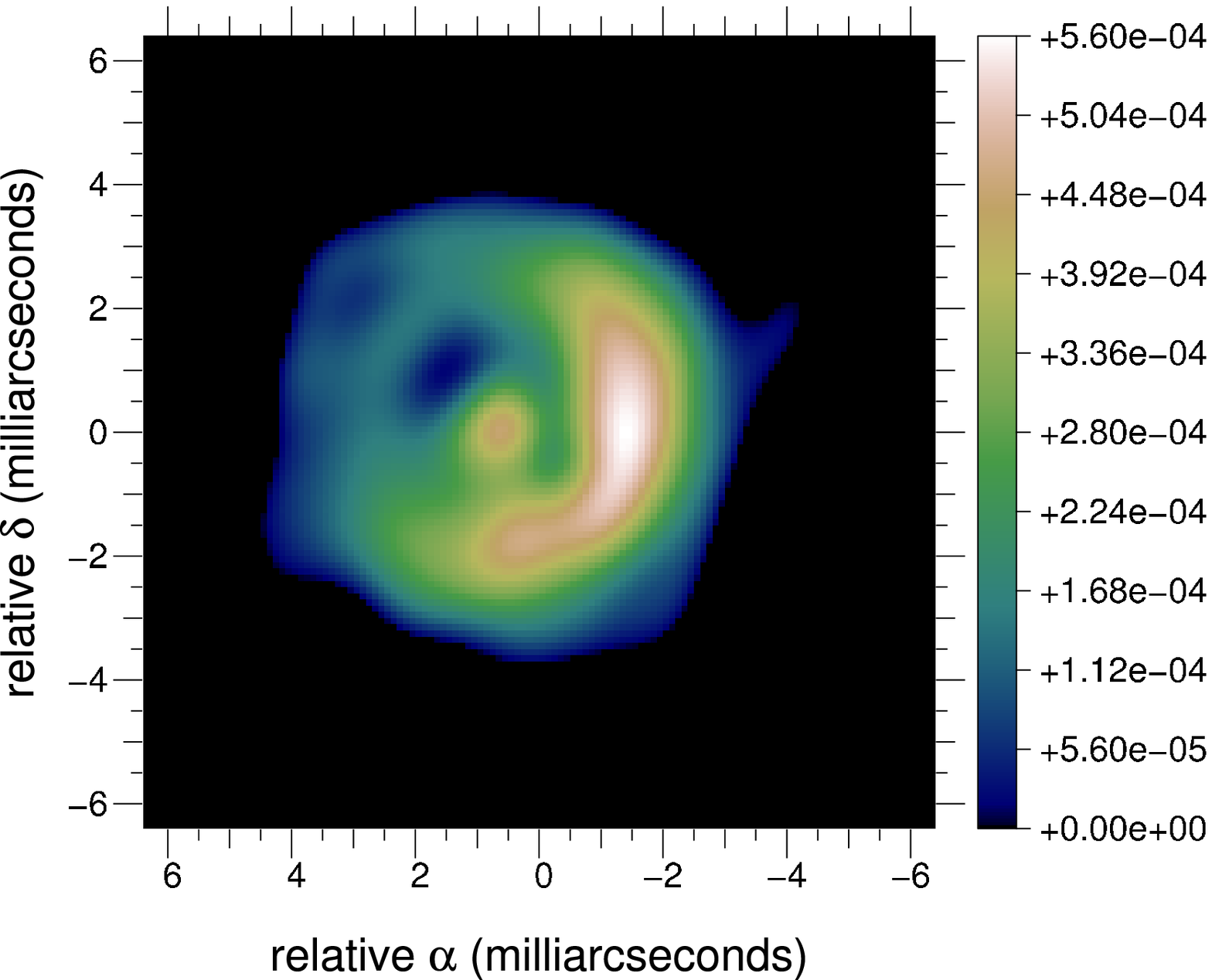}
  \includegraphics[height=33mm]{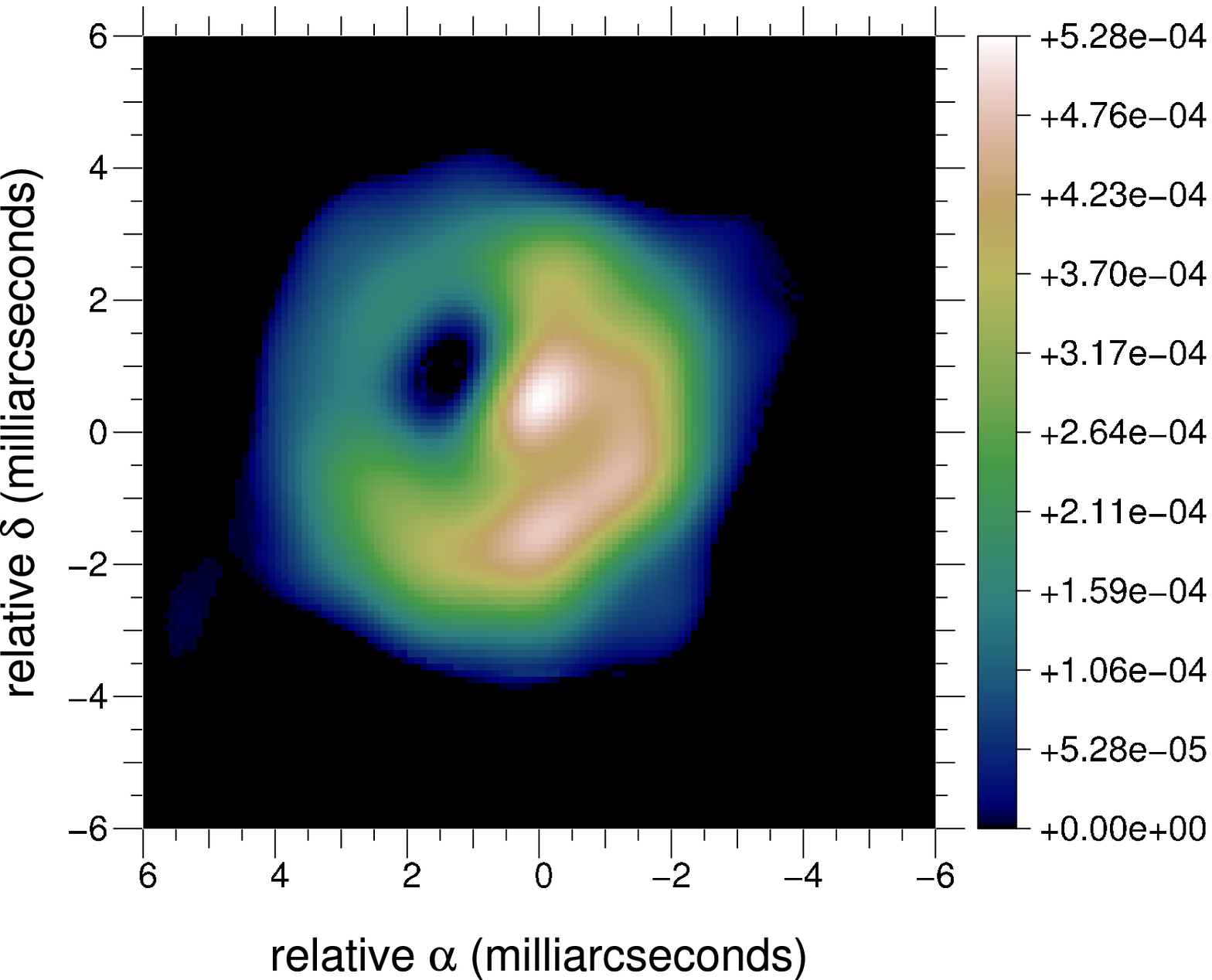}
  \caption{Image reconstruction with phase closure (left) and without any
    Fourier phase information (right).}
  \label{fig:no-phase}
\end{figure}

\textit{1) BSMEM} algorithm \cite{Buscher-1994-BSMEM,
  Baron_Young-2008-Marseille} makes use of a Maximum Entropy Method (\cf
Section~\ref{sec:MEM}) to regularize the problem of image restoration from the
measured bispectrum (hence its name).  The improved BSMEM version
\cite{Baron_Young-2008-Marseille} uses the Gull and Skilling entropy, see
\Eq{eq:mem-prior}, and a likelihood term with respect to the complex
bispectrum which assumes independent Gaussian noise statistics for the
amplitude and phase of the measured bispectrum.  The optimization engine is
\textsc{MemSys} which implements the strategy proposed by Skilling \& Bryan
\cite{Skilling_Bryan-1984-maximum_entropy} and automatically finds the most
likely value for the hyper-parameter $\mu$.  The default image is either a
Gaussian, a uniform disk, or a Dirac centered in the field of view.  Because
it makes no attempt to directly convert the data into complex visibilities, a
strength of \BSMEM is that it can handle any type of data sparsity (such as
missing closures).  Thus, in principle, \BSMEM could be used to restore images
when Fourier phase data are completely missing (see Fig.~\ref{fig:no-phase}).

\textit{2) The Building Block Method}
\cite{Hofmann_Weigelt-1993-building_blocks} is similar to the \Clean method
but designed for reconstructing images from bispectrum data obtained by means
of speckle or long baseline interferometry.  The method proceeds iteratively
to reduce a cost function $\Fdata^\BispectrumTag$ equal to that in
\Eq{eq:bispectrum-penalty} with weights set to a constant or to an expression
motivated by Wiener filtering.  The minimization of the penalty is achieved by
a matching pursuit algorithm which imposes sparsity of the solution.  The
image is given by the building block model in
Eq.'s~(\ref{eq:general-image-model})-(\ref{eq:grid-image-model}) and, at the
$n$-th iteration, the new image $\Image^{[n]}(\VDirn)$ is obtained by adding a
new building block at location $\VDirn^{[n]}$ with a weight $\alpha^{[n]}$ to
the previous image, so as to maintain the normalization:
\begin{displaymath}
  %\label{eq:building-block-next-image-normalized}
  \Image^{[n]}(\VDirn)
   = (1 - \alpha^{[n]})\,\Image^{[n - 1]}(\VDirn)
  + \alpha^{[n]} \, \BasisFunc(\VDirn - \VDirn^{[n]}) \, .\notag
\end{displaymath}
The weight and location of the new building block is derived by minimizing the
criterion $\Fdata^\BispectrumTag$ \wrt these parameters.  Strict positivity
and support constraint can be trivially enforced by limiting the possible
values for $\alpha^{[n]}$ and $\VDirn^{[n]}$.  To improve the convergence, the
method allows to add/remove more than one block at a time.  To avoid super
resolution artifacts, the final image is convolved with a smoothing function
with size set according to the spatial resolution of the instrument.

%\begin{align}
%  %\label{eq:building-block-next-image}
%  \Image^{[n]}(\VDirn)
%  & = \Image^{[n - 1]}(\VDirn)
%     + \alpha^{[n]} \, \BasisFunc(\VDirn - \VDirn^{[n]}) \notag\\
%\intertext{or, if strict normalization is applied:}
%  %\label{eq:building-block-next-image-normalized}
%  \Image^{[n]}(\VDirn)
%  & = (1 - \alpha^{[n]})\,\Image^{[n - 1]}(\VDirn)
%  + \alpha^{[n]} \, \BasisFunc(\VDirn - \VDirn^{[n]}) \, .\notag
%\end{align}

\textit{3) MACIM algorithm} \cite{Ireland_et_al-2006-MACIM}, for MArkov Chain
IMager, aims at maximizing the posterior probability:
\begin{displaymath}
  \Pr(\VParam\vert\Data)\propto
  \exp\Bigl(-\frac{1}{2}\,\Fdata(\VParam)
  - \frac{\mu}{2}\,\Fprior(\VParam)\Bigr) \,.
\end{displaymath}
MACIM implements MEM regularization and a specific regularizer which favors
large regions of dark space in-between bright regions.  For this latter
regularization, $\Fprior(\VParam)$ is the sum of all pixels with zero flux on
either side of their boundaries.  MACIM attempts to maximize
$\Pr(\VParam\vert\Data)$ by a simulated annealing algorithm with the
Metropolis sampler.  Although maximizing $\Pr(\VParam\vert\Data)$ is the same
as minimizing $\Fdata(\VParam)+\mu\,\Fprior(\VParam)$, the use of
\emph{normalized} probabilities is required by the Metropolis sampler to
accept or reject the image samples.  In principle, simulated annealing is able
to solve the global optimization problem of maximizing
$\Pr(\VParam\vert\Data)$ but the convergence of this kind of Monte-Carlo
method for such a large problem is very slow and critically depends on the
parameters which define the temperature reduction law.  A strict Bayesian
approach can also be exploited to derive, in a statistical sense, the values
of the hyper-parameters (such as $\mu$) and some \emph{a posteriori}
information such as the significance level of the image.

\textit{4) \Mira algorithm} \cite{Thiebaut-2008-Marseille} defines the sought
image as the minimum of the penalty function in Eq.~(\ref{eq:cost-function}).
Minimization is done by a limited variable memory method (based on BFGS
updates) with bound constraints for the positivity
\cite{Thiebaut:spie2002:bdec}.  Since this method does not implement global
optimization, the image restored by \Mira depends on the initial image.  \Mira
is written in a modular way: any type of data can be taken into account by
providing a function that computes the corresponding penalty and its gradient.
For the moment, \Mira handles complex visibility, power spectrum and
closure-phase data via penalty terms given by \Eq{eq:fdata-local-approx},
\Eq{eq:powerspectrum-penalty} and \Eq{eq:Mira-phase-closure-penalty}.  Also
many different regularizers are built into \Mira (negentropy, quadratic or
edge-preserving smoothness, compactness, total variation, \etc) and provisions
are made to implement custom priors.  \Mira can cope with any missing data, in
particular, it can be used to restore an image given only the power spectrum
(\ie without any Fourier phase information) with at least a $180^\circ$
orientation ambiguity.  An example of reconstruction with no phase data is
shown in Fig.~\ref{fig:no-phase}.  In the case of non-sparse \uv coverage, the
problem of image reconstruction from the modulus of its Fourier transform has
been addressed by Fienup~\cite{Fienup-1978} by means of an algorithm based on
projections onto convex sets (POCS).

%To compute the model of the complex visibilities from the current image, \Mira
%uses the \emph{exact} linear transform, \Eq{eq:matrix-coef-grid}, or Fourier
%interpolation (\cf Sect.~\ref{sec:Fourier-interpolation}) to speed up
%computations or to deal with large data sets.

\textit{5) Wisard algorithm} \cite{Meimon_et_al-2005-weak_phase_imaging}
recovers an image from power spectrum and phase closure data.  It exploits a
self-calibration approach (\cf Section~\ref{sec:self-calibration}) to recover
missing Fourier phases.  Given a current estimate of the image and the phase
closure data, \Wisard first derives missing Fourier phase information in such
a way as to minimize the number of unknowns.  Then, the synthesized Fourier
phases are combined with the square root of the measured power spectrum to
generate pseudo complex visibility data which are fitted by the image
restoration step.  This step is performed by using the chosen regularization
and a penalty with respect to the pseudo complex visibility data.  However, to
account for a more realistic approximation of the distribution of complex
visibility errors, \Wisard make uses of a quadratic penalty which is different
from the usual Goodman approximation
\cite{Meimon_et_al-2005-convex_approximation}.  Taken separately, the image
restoration step is a convex problem with a unique solution, the
self-calibration step is not strictly convex but (like in original
self-calibration method) does not seem to pose insurmountable problems.
Nevertheless, the global problem is multi-modal and, at least in difficult
cases, the final solution depends on the initial guess.  There are many
possible regularizers built into \Wisard, such as the one in
\Eq{eq:fov-penalty} and the \emph{edge-preserving smoothness} prior in
\Eq{eq:edge-preserving-roughness}.

\Mira and \Wisard have been developed in parallel and share some common
features.  They use the same optimization engine \cite{Thiebaut:spie2002:bdec}
and means to impose positivity and normalization
\cite{LeBesnerais_et_al-2008-interferometry}.  They however differ in the way
missing data is taken into account: \Wisard takes a self-calibration (\cf
Section~\ref{sec:self-calibration}) approach to \emph{explicitly} solve for
missing Fourier phase information; while \Mira \emph{implicitly} accounts for
any lack of information through the direct model of the
data\cite{LeBesnerais_et_al-2008-interferometry}.

%\subsection{Comparison of algorithms on simulated data}

All these algorithms have been compared on simulated data during
\emph{Interferometric Beauty Contests}
\cite{Lawson_etal-2004-image_beauty_contest,
  Lawson_etal-2006-image_beauty_contest, Cotton_et_al-2008-Marseille}.  The
results of the contest were very encouraging. Although being quite different
algorithms, \BSMEM, the Building Block Method, \Mira and \Wisard give good
image reconstructions where the main features of the objects of interest can
be identified in spite of the sparse \uv coverage, the lack of some Fourier
phase information and the non-linearities of the measurements.  \BSMEM and
\Mira appear to be the most successful algorithms (they respectively won the
first two and last contests).

\begin{figure}[!t]
  \centering
  \begin{tabular}{lr}
  \includegraphics[height=33mm]{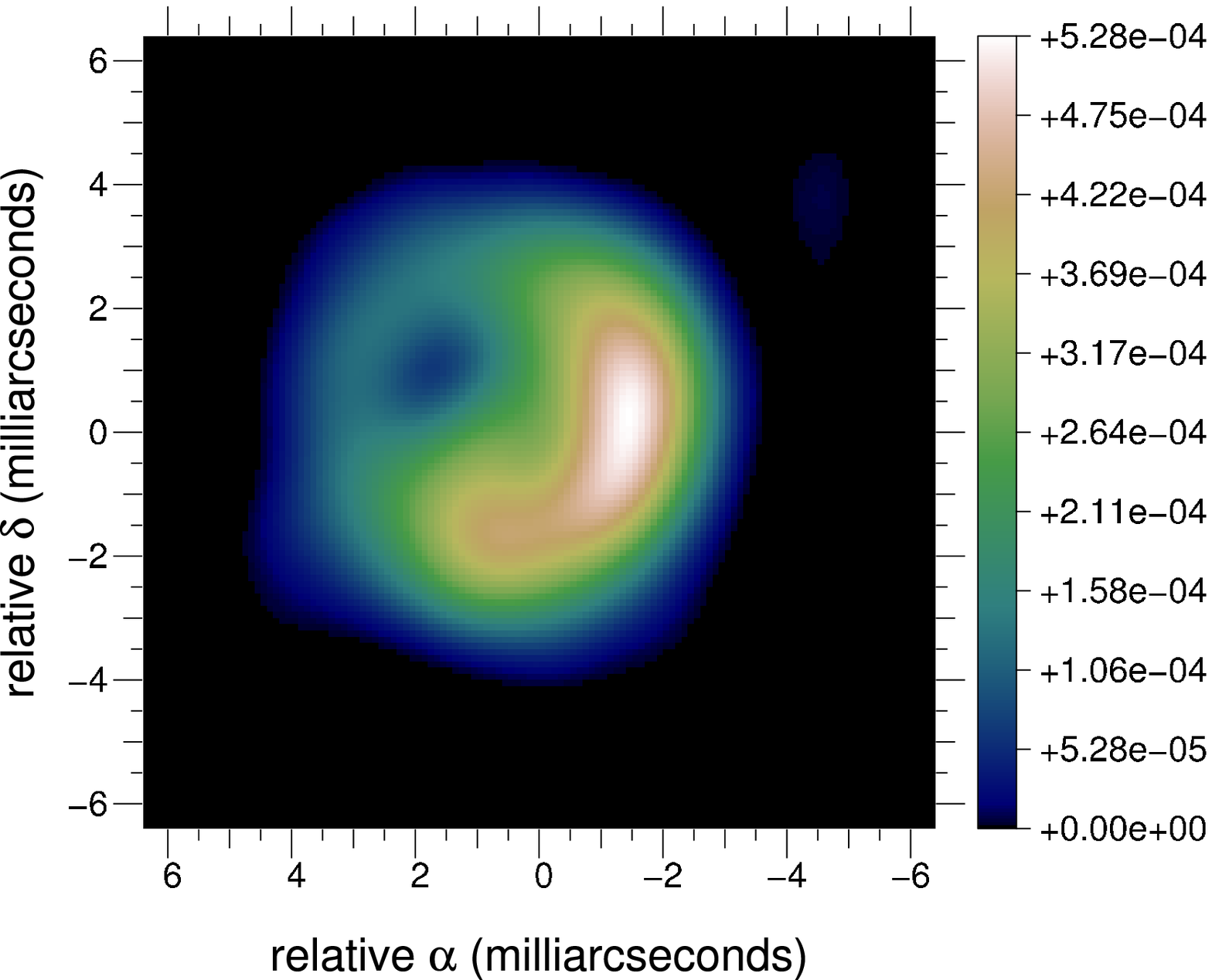} &
  \includegraphics[height=33mm]{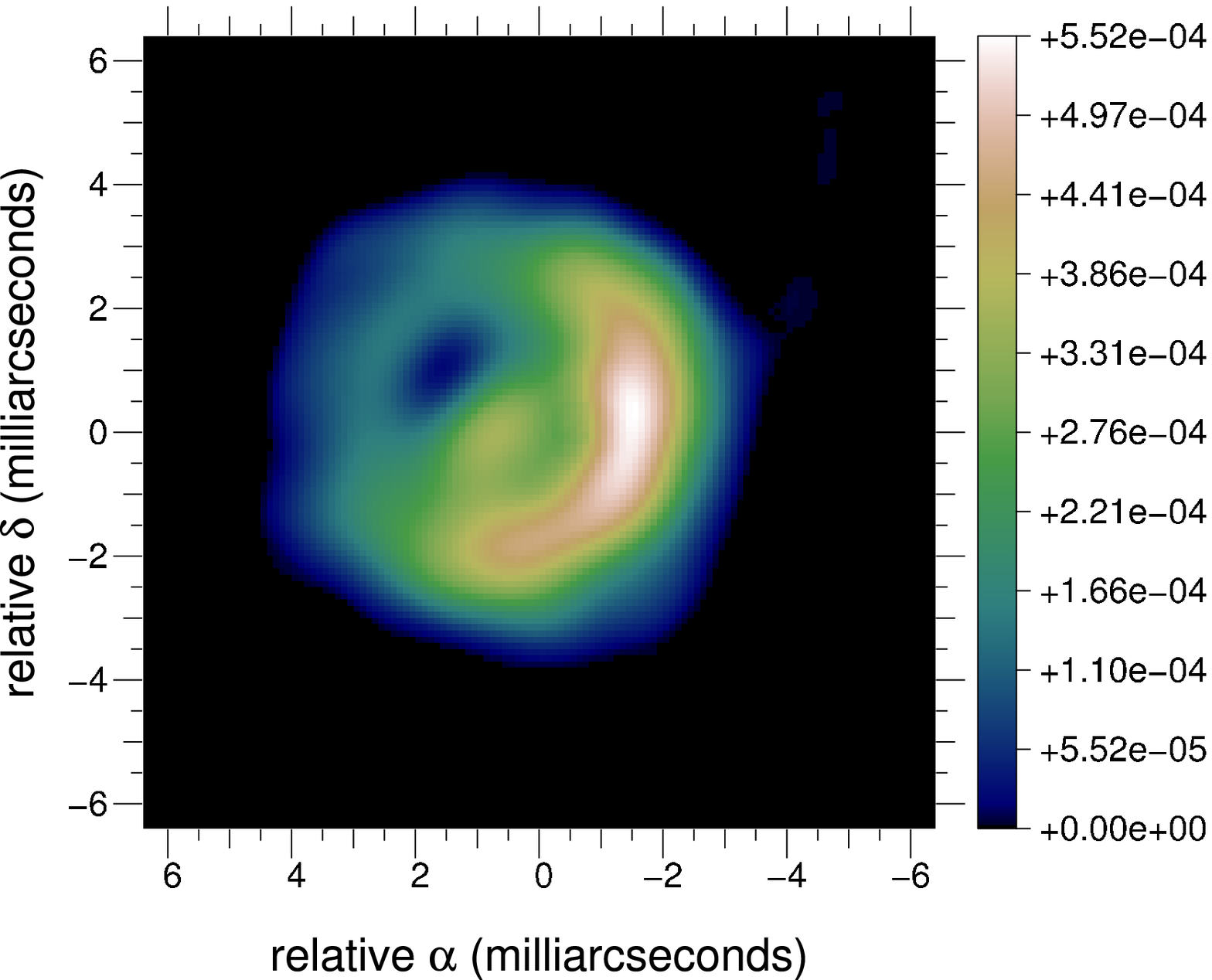} \\[2mm]
  \includegraphics[height=33mm]{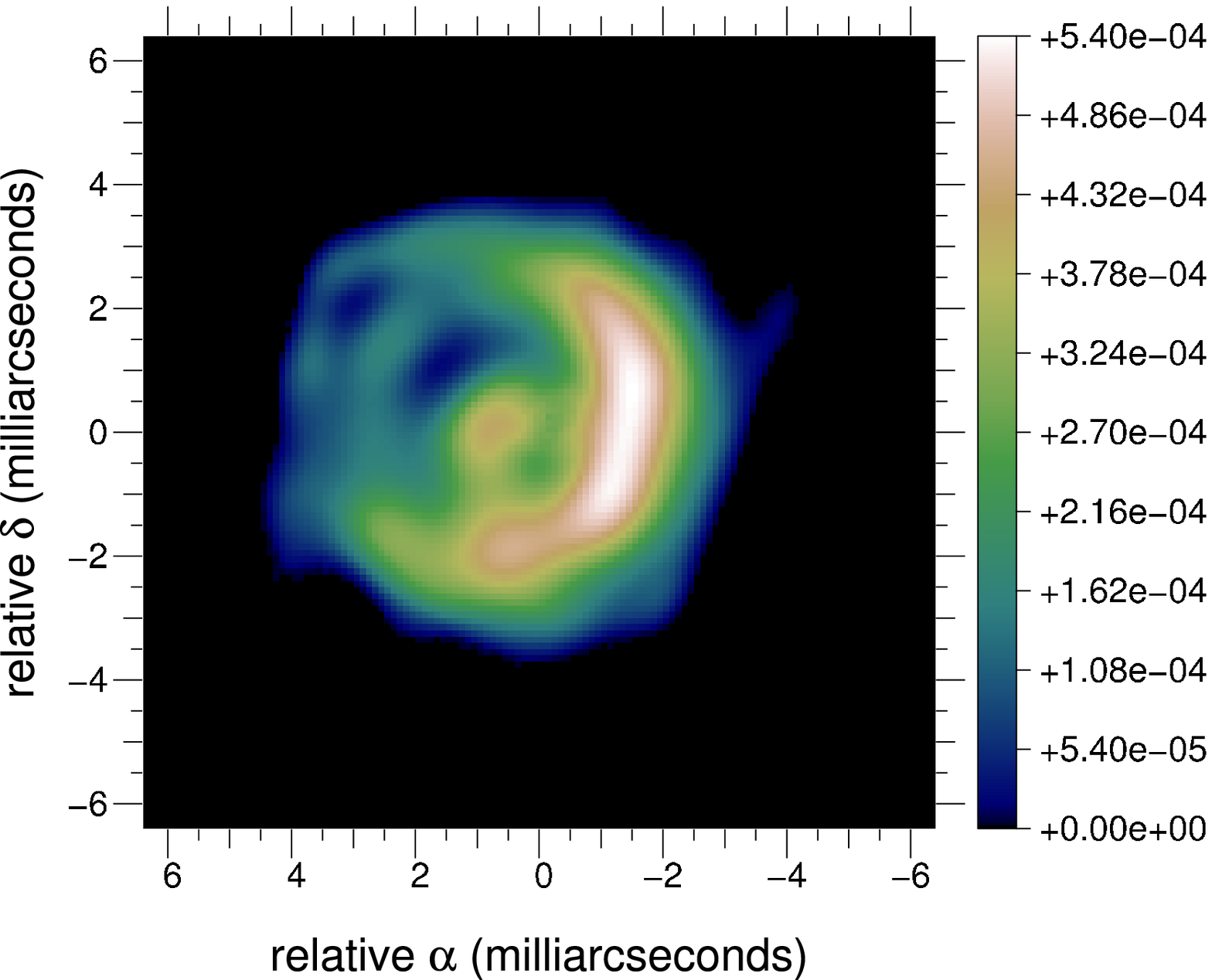} &
  \includegraphics[height=33mm]{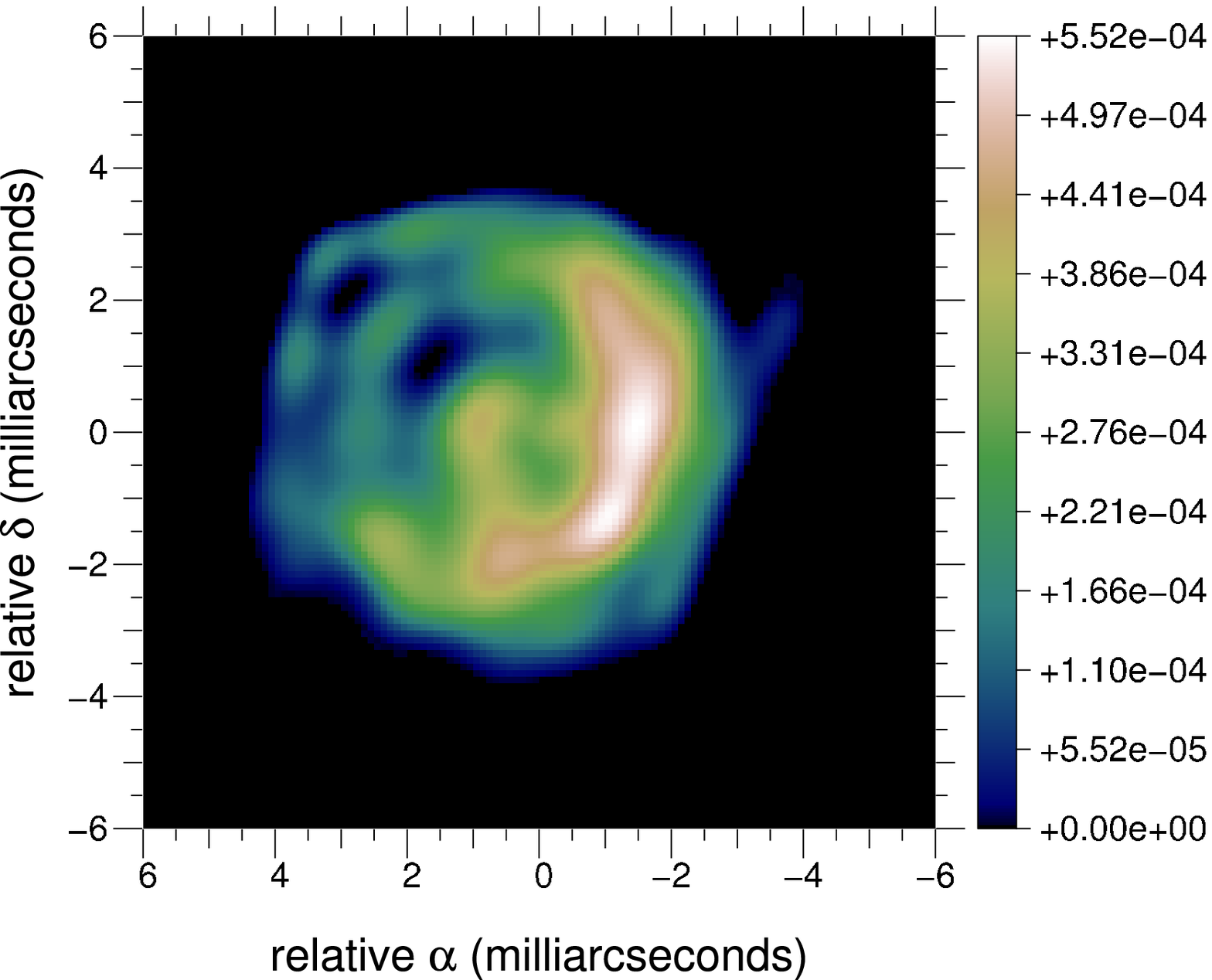}\\
  \end{tabular}
  \caption{Image reconstruction under various regularization levels.
    Algorithm is \Mira with edge-preserving regularization given in
    \Eq{eq:edge-preserving-roughness} with $\epsilon=10^{-4}$ and $\mu=10^{6}$
    (top-left), $\mu=10^{5}$ (top-right), $\mu=10^{4}$ (bottom-left) and
    $\mu=3\times10^{3}$ (bottom-right).}
  \label{fig:regularization-levels}
\end{figure}

With their tuning parameters and, for some of them, the requirement to start
with an initial image, these algorithms still need some expertise to be used
successfully.  But this is quite manageable.  For instance, the tuning of the
regularization level can be derived from Bayesian considerations but can also
almost be done by visual inspection of the restored image.  From
Fig.~\ref{fig:regularization-levels}, one can see the effects of
under-regularization (which yields more artifacts) and over-regularization
(which yields over simplification of the image).  In that case, a good
regularization level is probably between $\mu=10^{5}$ and $\mu=10^{4}$ and any
choice in this range would give a \emph{good} image.
Figure~\ref{fig:regularization-types} shows image reconstructions from one of
the data sets of the \emph{2004' Beauty Contest}
\cite{Lawson_etal-2004-image_beauty_contest} and with different types of
regularization.  These synthesized images do not greatly differ and are all
quite acceptable approximations of the reality (compare for instance with the
dirty image in Fig.~\ref{fig:dirty-map}).  Hence, provided that the level of
the priors is correctly set, the particular choice of a given regularizer can
be seen as a refinement that can be done after some reconstruction attempts
with a prior that is simpler to tune.  At least, the qualitative type of prior
is what really matters, not the specific expression of the penalty imposing
the prior.

%==============================================================================
%================================================================= DISCUSSION =
%==============================================================================

\section{Discussion}

The main issues in image reconstruction from interferometric data are the
sparsity of the measurements (which sample the Fourier transform of the object
brightness distribution) and the lack of part of the Fourier phase
information.  The inverse problem approach appears to be suitable to describe
the most important existing algorithms in this context.  Indeed, the image
reconstruction methods can be stated as the minimization of a mixed criterion
under some strict constraints such as positivity and normalization.  Two
different types of terms appear into this criterion: likelihood terms which
enforce consistency of the model image with the data, and regularization terms
which maintain the image close to the priors required to lever the
degeneracies of the image reconstruction problem.  Hence, the differences
between the various algorithms lie in the kind of measurements considered, in
the approximations for the direct model and for the statistics of the errors
and in the prior imposed by the regularization.  For non-convex criteria which
occur when the OTF is unknown or when non-linear estimators are measured to
overcome turbulence effects, the initial solution and the optimization
strategy are also key components of the algorithms.  Although \emph{global
  optimization} is required to solve such multi-modal problems, most existing
algorithms are successful whereas they only implement local optimization.
These algorithms are not fully automated black boxes: at least some tuning
parameters and the type of regularization are left to the user choice.
Available methods are however now ready for image reconstruction from
\emph{real} data.  Nevertheless, a general understanding of the mechanisms
involved in image restoration algorithms is mandatory to correctly use these
methods and to analyze possible artifacts in the synthesized images.  From a
technical point of view, future developments of these algorithms will
certainly focus on global optimization and unsupervised reconstruction.
However, to fully exploit the existing instruments, the most worthwhile tracks
to investigate are multi-spectral imaging and accounting for additional data
such as a low resolution image of the observed object to overcome the lack of
short baselines.

%\section*{Acknowledgment}
% 
%The authors are grateful to two anonymous reviewers for their constructive and
%fruitful comments.

%==============================================================================
%=============================================================== BIBLIOGRAPHY =
%==============================================================================

\newcommand{\aj}{Astron. J.} % Astronomical Journal
\newcommand{\araa}{Annual Rev. of Astron. \& Astrophys.} % Annual Review of Astron and Astrophys
\newcommand{\apj}{Astrophys. J.} % Astrophysical Journal
\newcommand{\apjl}{Astrophys. J. Lett.} % Astrophysical Journal, Letters
\newcommand{\apjs}{Astrophys. J. Suppl.} % Astrophysical Journal, Supplement
\newcommand{\ao}{Appl. Optics} % Applied Optics
\newcommand{\optlett}{Optics Lett.} % Optics Letter
\newcommand{\josa}{J.\ Opt.\ Soc.\ Am.} % Journal of the Optical Society of America
\newcommand{\josaa}{J.\ Opt.\ Soc.\ Am.\ A} % Journal of the Optical Society of America A 
\newcommand{\josab}{J.\ Opt.\ Soc.\ Am.\ B} % Journal of the Optical Society of America B 
\newcommand{\apss}{Astrophys. \& Space Science} % Astrophysics and Space Science
\newcommand{\aap}{Astron. \& Astrophys.} % Astronomy and Astrophysics
\newcommand{\aapr}{Astron. \&  Astrophys. Rev.} % Astronomy and Astrophysics Reviews
\newcommand{\aaps}{Astron. \& Astrophys. Suppl.} % Astronomy and Astrophysics, Supplement
\newcommand{\azh}{Astronomicheskii Zhurnal} % Astronomicheskii Zhurnal
\newcommand{\baas}{Bulletin of the AAS} % Bulletin of the AAS
\newcommand{\jrasc}{J. of the RAS of Canada} % Journal of the RAS of Canada
\newcommand{\memras}{Memoirs of the RAS} % Memoirs of the RAS
\newcommand{\mnras}{Monthly Notices of the RAS} % Monthly Notices of the Royal Astronomical Society
\newcommand{\pra}{Physical Rev. A: General Physics} % Physical Review A: General Physics
\newcommand{\prb}{Physical Rev. B: Solid State} % Physical Review B: Solid State
\newcommand{\prc}{Physical Rev. C} % Physical Review C
\newcommand{\prd}{Physical Rev. D} % Physical Review D
\newcommand{\pre}{Physical Rev. E} % Physical Review E
\newcommand{\prl}{Physical Rev. Lett.} % Physical Review Letters
\newcommand{\pasp}{Publications of the ASP} % Publications of the ASP
\newcommand{\pasj}{Publications of the ASJ} % Publications of the ASJ
\newcommand{\qjras}{Quarterly J. of the RAS} % Quarterly Journal of the RAS
\newcommand{\skytel}{Sky and Telescope} % Sky and Telescope
\newcommand{\solphys}{Solar Physics} % Solar Physics
\newcommand{\sovast}{Soviet Astronomy} % Soviet Astronomy
\newcommand{\ssr}{Space Science Rev.} % Space Science Reviews
\newcommand{\zap}{Zeitschrift fuer Astrophysik} % Zeitschrift fuer Astrophysik
\newcommand{\nat}{Nature} % Nature
\newcommand{\iaucirc}{IAU Cirulars} % IAU Cirulars
\newcommand{\aplett}{Astrophys. Lett.} % Astrophysics Letters
\newcommand{\apspr}{Astrophys. Space Physics Research} % Astrophysics Space Physics Research
\newcommand{\bain}{Bulletin Astronomical Institute of the Netherlands} % Bulletin Astronomical Institute of the Netherlands
\newcommand{\fcp}{Fundamental Cosmic Physics} % Fundamental Cosmic Physics
\newcommand{\gca}{Geochimica Cosmochimica Acta} % Geochimica Cosmochimica Acta
\newcommand{\grl}{Geophysics Research Lett.} % Geophysics Research Letters
\newcommand{\jcp}{J. of Chemical Physics} % Journal of Chemical Physics
\newcommand{\jgr}{J. of Geophysics Research} % Journal of Geophysics Research
\newcommand{\jqsrt}{J. of Quantitiative Spectroscopy and Radiative Transfer} % Journal of Quantitiative Spectroscopy and Radiative Transfer
\newcommand{\memsai}{Mem. Societa Astronomica Italiana} % Mem. Societa Astronomica Italiana
\newcommand{\nphysa}{Nuclear Physics A} % Nuclear Physics A
\newcommand{\physrep}{Physics Reports} % Physics Reports
\newcommand{\physscr}{Physica Scripta} % Physica Scripta
\newcommand{\planss}{Planetary Space Science} % Planetary Space Science
\newcommand{\procspie}{Proc. SPIE} % Proceedings of the SPIE

%\bibliographystyle{IEEEtran}
%\bibliography{IEEEabrv,spm09}

%==============================================================================
%================================================================ BIOGRAPHIES =
%==============================================================================

\footnotesize%\baselineskip=8pt\renewcommand{\baselinestretch}{0.8}%

\medskip

\noindent\textbf{Eric Thi\'ebaut} was born in B\'eziers, France, in 1966.  He
graduated from the {\'E}cole Normale Sup{\'e}rieure in 1987 and received the
Ph. D. degree in Astrophysics at Universit{\'e} Pierre \& Marie Curie (Paris
VII, France), in 1994.  Since 1995, he is an astronomer at the Centre de
Recherche Astrophysique de Lyon.  His main interests are in the fields of
signal processing and image reconstruction. He made various contributions in
blind deconvolution, optical interferometry and optimal detection with
applications in astronomy, bio-medical imaging and digital holography.

\medskip

\noindent\textbf{Jean-Fran\c{c}ois Giovannelli} was born in B\'eziers, France,
in 1966.  He graduated from the \'Ecole Nationale Sup\'erieure de
l'\'Electronique et de ses Applications in 1990. He received the Ph. D. degree
in 1995 and the \emph{Habilitation \`a Diriger des Recherches} in physics
(signal processing) in 2005.  From 1997 to 2008, he has been Assistant
Professor with the Universit\'e Paris-Sud, and a Researcher with the
Laboratoire des Signaux et Syst\`emes.  He is presently Professor with the
Universit\'e de Bordeaux and a Researcher with the Laboratoire d'Int\'egration
du Mat\'eriau au Syst\`eme, \'Equipe Signal-Image.  He is interested in
regularization and Bayesian methods for inverse problems in signal and image
processing. His application fields essentially concern astronomical, medical,
protemics and geophysical imaging.

\vfill \vbox{}

\end{document}